\documentclass[11pt]{article}

\usepackage{mathrsfs,amssymb,amstext,amsmath,amsthm,eucal}
\usepackage{graphicx,subfig}
\usepackage[usenames,dvipsnames]{color}
\usepackage{geometry}

\newcommand{\lie}{\mathcal{L}}
\newcommand{\kto}[3]{#1 \stackrel{{\pi}}{\mapsto} ( #2 ,\,#3 ) }
\newcommand{\AdS}{\ensuremath{\mathrm{AdS}_3} }
\newcommand{\sub}[3]{{}_{ {#1}_{#2} \cdots {#1}_{#3} }{} }
\newcommand{\subn}[2]{\sub{\nu}{#1}{#2}}
\newcommand{\grad}{\mathrm{d}}

\newcommand{\newadd}[1]{{#1}}

\usepackage{authblk}
\title{No more CKY two-forms in the NHEK}
\author{Yoshihiro Mitsuka\footnote{yoshihiro.mitsuka@gmail.com}}
\author{George Moutsopoulos\footnote{gmoutso@googlemail.com} }
\affil{Department of Electrophysics, National Chiao-Tung University, Hsinchu, Taiwan, R.O.C.}

\usepackage{hyperref}
\hypersetup{ pdftitle   = {No more CKY two-forms in the NHEK},  pdfkeywords = {conformal, Killing-Yano, symmetries, black hole, Kerr},  pdfauthor  = {Yoshihiro Mitsuka, George Moutsopoulos},  pdfcreator = {\LaTeX}}

\date{October 11, 2011}

\begin{document}
\maketitle
\begin{abstract}We show that in the near-horizon limit of a Kerr-NUT-AdS black hole, the space of conformal Killing-Yano two-forms does not enhance and remains of dimension two. The same holds for an analogous polar limit in the case of extremal NUT charge. We also derive the conformal Killing-Yano $p$-form equation for any background in arbitrary dimension in the form of parallel transport.\end{abstract}

\section{Introduction and Conclusion}
Killing-Yano (KY) forms  suitably generalize the notion of Killing vectors to higher-degree differential forms~\cite{Kashiwada:1968}. They are related to  constants of geodesic motion \cite{Krtous:2006qy}, symmetries of the Dirac equation \cite{Benn:1996ia}, exotic supersymmetries of the superparticle \cite{Santillan:2011sh}, ADM-like charges \cite{Kastor:2004jk}, and the integrability of the Hamilton-Jacobi equations and the Klein-Gordon equation \cite{Yasui:2011pr}. Two recent, complementary, reviews are given in \cite{Santillan:2011sh} and \cite{Yasui:2011pr}. Conformal Killing-Yano (CKY) forms are the conformal generalization  of Killing-Yano forms and can describe the symmetries of massless or conformally invariant equations. In four dimensions the only degree of CKY forms to study other than one is, by Hodge duality, the degree of two-forms.

Rasmussen shows in \cite{Rasmussen:2010rw} that the principal Killing-Yano two-form of the $d=4$ Kerr-NUT-(Anti)-de Sitter black hole has a smooth limit under the near-horizon spacetime limit. At the same time, the CKY equation can be written in the form of parallel transport under some connection $D$ on a bundle of p-forms \cite{Semmelmann:2002}. Then Geroch's result on spacetime limits and the holonomy of $D$ \cite{Geroch:1969ca} implies that the vector space of CKY two-forms cannot reduce its dimension under the near-horizon limit, in accordance to \cite{Rasmussen:2010rw}. The space of CKY two-forms could enhance though. It was this question we wanted to answer definitely.  We find that there are only two independent CKY two-forms in the near-horizon geometry: the Killing-Yano two-form and its Hodge dual in \cite{Rasmussen:2010rw}.

For the near-horizon geometry, there is an enhancement of isometries from rank two, the time translations and rotations, to rank four that form $\mathfrak{sl}(2,\mathbb{R})\oplus\mathbb{R}$. This is expected by construction of the limit that can be written as a limit of a finite diffeomorphism,
\[ g_0 = \lim_{\zeta\rightarrow0^+} e^{\frac{1}{\zeta}(\tau\partial_\tau-x\partial_x)} g~.\]
In this expression, an infinitesimal diffeomorphism of the form $\tau\partial_\tau-x\partial_x$, which generates the finite diffeomorphism, becomes in the limit $\zeta\rightarrow 0^+$ a Killing vector of $g_0$. Given that the Killing vectors enhance, we asked whether the same holds for the Killing-Yano two-forms, and more generally whether the CKY two-forms enhance. There are for instance relations between Killing-Yano forms of various degrees in the definition of differential structure,  e.g. see \cite{Semmelmann:2002}. Note though that the near-horizon limit is not the same as the BPS limits of black holes in~\cite{Chen:2006xh}, which are known to have extra structure\newadd{, e.g. see \cite{Kubiznak:2009ad}}.

A recent result is the classification of spacetimes under the existence of a closed CKY two-form in~\cite{Krtous:2008tb,Houri:2008ng,Houri:2007xz}. However, our motivation was more in lines with~\cite{Cariglia:2011yt}. In particular, we were interested in the construction and consequences of a non-trivial (graded) algebra of Killing vectors and CKY two-forms. This could be achieved by studying the (graded) commutator of the symmetry operators on the Dirac equation~\cite{Benn:1996ia} as in \cite{Cariglia:2011yt} or by lifting them, in the case of special Killing-Yano, to parallel objects on the cone~\cite{Semmelmann:2002} as in~\cite{FigueroaO'Farrill:2008ka}. In the context of recent interest in the near-horizon geometry~\cite{Bredberg:2011hp}, but undoubtedly beyond that, extra structure or symmetries in the near-horizon geometry would be of considerable interest.

We also derive explicitly the CKY transport equation of $p$-forms, which was described in~\cite{Semmelmann:2002}. Here we were tempted to solve the equation explicitly. However, the calculation would have been quite involved and not illuminating. Since the derivation of the transport equation itself is straightforward but intense, we give the result in appendix~\ref{app:cky}. Instead, we solve the problem at hand by making use of the symmetry $\mathfrak{sl}(2,\mathbb{R})\oplus\mathbb{R}$ that renders the background of cohomogeneity one. This allows us to write an Ansatz on both sides of the CKY two-form defining equation, which is shown to not pass the test.

In \S\ref{sec:CKYtwo} we introduce CKY two-forms. We show how for $d=4$ Einstein solutions each CKY two-form is mapped to a pair of Killing vectors. Furthermore, the map transforms equivariantly under the isometry algebra. In \S\ref{sec:KerrAdS} we introduce the Kerr-NUT-AdS black hole and the coordinate ranges that we use. In \S\ref{sec:nhek} we introduce the near-horizon limit of these black holes. In addition to \cite{Rasmussen:2010rw}, we discuss when the near-horizon limit has well-defined coordinate ranges, as inherited from the black hole. This happens only when the NUT charge is zero. We can thus differentiate the limit from a solution generating technique to a limit that can describe a physical process when there is no NUT charge. We also discuss in parallel another spacetime limit, which we dub the polar limit. It is similar to the near-horizon limit but with the role of radial and polar coordinates exchanged. We find that the polar limit has well-defined coordinate ranges for any non-zero NUT charge. In these two sections, \S\ref{sec:KerrAdS} and \S\ref{sec:nhek}, we briefly comment on the positive cosmological constant case. 

In \S\ref{sec:noparallel} we show that there are no parallel two-forms in these two limits because the Levi-Civita holonomy is not special. In \S\ref{sec:LargerIsEight} we show that the CKY two-form equation in the NHEK background can have either two or eight independent solutions. This follows because the CKY two-forms should transform under $\mathrm{SL}(2,\mathbb{R})$ and that there are no parallel two-forms. If there are eight independent solutions, we can write an Ansatz for the ``new'' CKY two-forms by using the symmetries of the metric. However the Ansatz does not pass the test of the CKY equation as we show in the concluding section~\S\ref{sec:Ansatz}. 

The appendices contain supplementary material for our work. Appendix~\ref{app:cky} derives the connection $D$ under which CKY $p$-forms are parallel. Appendix~\ref{app:pureAdS} describes a diffeomorphism of AdS$_4$, which motivates the coordinate range of Anti-de Sitter black holes with NUT charge. Appendix~\ref{app:dSProfiles} repeats the analysis of section~\ref{sec:KerrAdS} for positive cosmological constant. Appendix~\ref{app:AdS3} is a brief classification of AdS$_3$ coordinates and appendix~\ref{app:solvingNHEK} solves the Einstein equations for a specific form of the metric.

\section{Conformal Killling-Yano two-forms}\label{sec:CKYtwo}
A two-form $K$ on a $d$-dimensional manifold $M$ with metric $g$ is a conformal Killing-Yano two-form (CKY) if it satisfies
\begin{equation}\label{eq:CKY2eqInd}
 \nabla_\mu K_{\nu\rho} = A_{\mu\nu\rho} +\frac{1}{2} g_{\mu\nu}B_\rho - \frac{1}{2} g_{\mu\rho}B_\nu ~,
\end{equation}
where $\nabla$ is the Levi-Civita derivative, $A$ is a three-form and $B$ a one-form. A CKY two-form with $B=0$ is called a Killing-Yano two-form. CKY two-forms are in one-to-one correspondence with $D$-parallel sections of $\mathcal{E}$,
\begin{equation}
K \stackrel{=}{\mapsto} E= K + A + B + C \in \Lambda^2 \oplus \Lambda^{3}\oplus \Lambda^{1}\oplus \Lambda^2=\mathcal{E}~,
\end{equation}
where $\Lambda^p$ the space of $p$-forms and $D$ a connection on $\mathcal{E}$. In appendix~\ref{app:cky} we derive the connection $D$ for the general case of a CKY $p$-form\footnote{\newadd{For $p=1$, these are the (metric dual of) conformal Killing vectors. Nevertheless, for $p>1$ one cannot in general associate to the $p$-form a derivation generalizing the Lie derivative.}}.

The transport equation was described in \cite{Semmelmann:2002}, the calculation tool prefered instead being B\"ar's cone construction.  However, the latter construction is possible only for the so-called special CKY $p$-forms. The first-order equations $D_\mu E = 0$ can be, in principle, solved up to the obstructions given by the holonomy of $D$. The transport equations have rank $d(d+1)(d+2)/6$. For instance, in $d=4$ there are $20$ first-order equations and solving them for a background of interest is computationally involved. 

In the case of an Einstein space of dimension $d=4$, we will use the simplification that both $B$ and the Hodge dual of the three-form $A$ in \eqref{eq:CKY2eqInd} are Killing one-forms, as noted already in~\cite{Jezierski:2005cg}. The connection $D$ is then given by \eqref{eq:CKY2eqInd} and
\begin{align}
\nabla_\mu A_{\nu_1\nu_2\nu_3} & {=} 
-\frac{3}{2} R_{[\nu_1\nu_2 |\mu}{}^\sigma K_{\sigma|\nu_3]} - \frac{3}{4} g_{\mu[\nu_1} C_{\nu_2\nu_3]} \\
\nabla_\mu B_\nu & = \frac{1}{2} C_{\mu\nu} \\
\nabla_\mu C_{\nu_1\nu_2} & =
-2 R_{\nu_1\nu_2\mu}{}^\sigma B_\sigma\label{eq:Killings}~.
\end{align}
The Levi-Civita derivative on the one-form $B$, $\nabla_\mu B_\nu$, is antisymmetric in its indices and thus, by definition, $B_\mu$ is a Killing one-form. This result was already shown by Tachibana in~\cite{Tachibana:1968}. The content of \eqref{eq:Killings} is not more than Killing's identity for a Killing vector. Furthermore, if $K$ satisfies
\begin{equation}\label{eq:CKY2eq}
 \nabla_X K = i_X A + \frac{1}{2} X^\flat \wedge B ~
\end{equation}
as in \eqref{eq:CKY2eqInd}, then its Hodge dual\footnote{
The square of the hodge dual in $d=4$ lorentzian signature is $\ast^2|_{\Lambda^2}=-1$ and $\ast^2|_{\Lambda^{1}\oplus\Lambda^3}=+1$. With indices these relations are 
 $1/(2!)^2 \, \epsilon_{\mu\nu}{}^{\rho\sigma}\epsilon_{\rho\sigma}{}^{\tilde\mu\tilde\nu}=-\delta^{[\tilde\mu}_\mu\delta^{\tilde\nu]}_\nu$, 
 $1/3!\, \epsilon_\mu{}^{\nu_1\cdots\nu_3}\epsilon_{\nu_1\cdots\nu_3}{}^{\tilde\mu}=\delta^{\tilde{\mu}}_\mu$
 and 
 $1/3!\, \epsilon_{\mu\nu\rho}{}^\sigma\epsilon_\sigma{}^{\tilde\mu\tilde\nu\tilde\rho}=\delta^{[\tilde\mu}_\mu\delta^{\tilde\nu}_\nu\delta^{\tilde\rho]}_\rho$,
 where $\epsilon_{\mu\nu\rho\sigma}$ is the Levi-Civita tensor.
} is again a CKY two-form and satisfies
\begin{equation}\label{eq:dualCKY2eq}
 \nabla_X \ast {K} = -\frac{1}{2} i_X \ast B + X^\flat \wedge \ast A ~.
\end{equation}
By the same token as we used for $B$, $\nabla_\mu (\ast A)_\nu$ is antisymmetric in its indices and is also a Killing one-form. It is this simplification that we will use in the present paper.

From the above, the CKY two-form $K$ on a four-dimensional Einstein space $(M,g)$ is mapped to a pair of Killing vectors. We will write this as $\kto{K}{\tilde\xi}{\xi}$ with $A=\ast\, g(\tilde\xi,-)$ and $B=\frac{1}{2} g(\xi,-)$. Note that the kernel of this map is given by those CKY two-forms that are parallel. It is easy to show, using \eqref{eq:dualCKY2eq}, that
if $\kto{K}{\tilde\xi}{\xi}$, then its Hodge dual is a CKY two-form with $\kto{\ast K}{-\xi}{\tilde\xi}$. That is, the complex structure of CKY two-forms $i\,K :=\ast K$ is compatible with the complex structure of the Killing vector doublets $i(\tilde\xi,\xi):=(-\xi,\tilde\xi)$. 

Knowledge of the isometries of the metric simplifies the task of finding the CKY two-forms, since the unknowns on the right-hand side of \eqref{eq:CKY2eq} are now in terms of a finite number of constants, a linear combination of the known Killing vectors. Killing vectors are easier to find in general, whereas their maximal rank is $d(d+1)/2=12$. However, the problem can be reduced further. The Lie derivative along a Killing vector $k$ commutes with the Hodge operator, and its commutator with the Levi-Cevita derivative satisfies $[\lie_k,\nabla_X]=\nabla_{[k,X]}$ for all vectors $X$. By using \eqref{eq:CKY2eq} we
can show that if $K$ is a CKY two-form with $\kto{K}{\tilde\xi}{\xi}$, then its Lie derivative along a Killing vector $k$ is also a CKY two-form with $\kto{\lie_k K}{ [k,\tilde\xi]}{ [k,\xi]}$,
\begin{equation}
 \begin{aligned}
 0 = &\lie_k\left( \nabla_X K - i_X A - \frac{1}{2} X^\flat \wedge B \right)\\
   =  &\nabla_{[k,X]} K - i_{[k,X]} A - \frac{1}{2} [k,X]^\flat \wedge B \\
  &  + \nabla_X  \lie_k K - i_X \lie_k A - \frac{1}{2} X^\flat \wedge \lie_k B \\
  = & \nabla_X  \lie_k K - i_X \lie_k A - \frac{1}{2} X^\flat \wedge \lie_k B~.
\end{aligned}
\end{equation}
Therefore, CKY two-forms form a representation under the isometry algebra of the metric and the map $\kto{K}{\tilde\xi}{\xi}$ is equivariant under the action of $\lie_k$. We will use the power of this result in what follows.

\section{Kerr (Nut) Anti-de Sitter black holes}\label{sec:KerrAdS}
Kerr's black holes in anti-de Sitter space with cosmological constant $-3/\ell^2$ is described by their mass $M$, a rotational parameter $a$ and the NUT charge $L$. The metric is given by
\begin{multline}\label{eq:KerrAdS1}
 \grad s^2  = - \frac{\Delta_r}{r^2+y^2} \left(  \grad \hat{t} + y^2  \grad \psi\right)^2
           + \frac{\Delta_y}{r^2+y^2} \left(  \grad \hat{t} - r^2  \grad \psi\right)^2\\
+\frac{r^2+y^2}{\Delta_r}  \grad r^2+\frac{r^2+y^2}{\Delta_y}  \grad y^2~,
\end{multline}
where the metric functions are
\begin{subequations}\label{eq:DeltaAdS}
\begin{align}
 \Delta_r &= (1+ \frac{r^2}{\ell^2})(r^2+a^2) - 2 M r~,\\
 \Delta_y &= (a^2-y^2)(1-\frac{y^2}{\ell^2})+2 L y~. 
\end{align}
\end{subequations}
These metrics can be generalized to the Ple\-ba\'n\-ski\--De\-mia\'n\-ski~\cite{Plebanski:1976gy} family of type D Ein\-stein\--Max\-well solutions\footnote{See~\cite{Kubiznak:2007kh}, which discusses their CKY two-forms.}. By using the symmetry of Chen, Lu and Pope from~\cite{Chen:2006xh}, which inverts $a/\ell\mapsto \ell/a$, we will always take $0\leq a\leq\ell$.

When $M=L=0$, the space is isometric to anti-de Sitter in two different ways, see appendix~\ref{app:pureAdS}. Each diffeomorphism corresponds to either the range $|y|<a$ or $|y|>\ell$, the two regions where the function $\Delta_y$ is positive. In particular, the range $|y|<a$ covers the whole of AdS$_4$. This implies we can focus on $|y|\leq a$ since the two ranges are isometric. As we add non-zero mass and NUT charge, we will continue to take $y$ in the finite region between the two middle roots of $\Delta_y$. With zero mass, $\Delta_r$ is positive for all $r$. Above some critical mass $M_*$ there are roots to $\Delta_r$ so that the singularities at $r=y=0$ are hidden from large distances by a horizon. The coordinate range of $(r,y)$ are thus determined from the roots of $\Delta_r$ and $\Delta_y$. One of our tasks is to give the different profiles of the graphs of these two functions as we vary $M$ and $L$ for fixed $a$ and $\ell$.

The value of NUT charge also affects the periodicity of the coordinates. When $L=0$, we shift $\hat{t}={t}-a^2\psi$ so that
\begin{multline}\label{eq:KerrAdS2}
 \grad s^2  = - \frac{\Delta_r}{r^2+y^2} \left(  \grad {t} + (y^2-a^2)  \grad \psi\right)^2
           + \frac{(a^2-y^2)(1-\frac{y^2}{\ell^2})}{r^2+y^2} \left(  \grad {t} - (r^2+a^2)  \grad \psi\right)^2 \\
+\frac{r^2+y^2}{\Delta_r}  \grad r^2+\frac{r^2+y^2}{(a^2-y^2)(1-\frac{y^2}{\ell^2})}  \grad y^2~.
\end{multline}
Smoothness close to $y=\pm a$ at constant $(t,r)$ requires $\psi$ to be periodic with
\begin{equation}
\psi=  \psi+\frac{2\pi}{a\left(1 - \frac{a^2}{\ell^2} \right)}~.
\end{equation}
We are also interested in how the periodicities change with $L\neq0$. 
As usual, a spacetime with NUT charge will require closed timelike curves.  In this section, we present the allowed coordinate ranges of $(r,y,t,\psi)$ for the various allowed choices of parameters $(a,M,L)$. 

\subsubsection*{Profiles}

\begin{figure}
  \centering
  \subfloat[$M<M_*$, no roots]{\includegraphics[width=0.3\textwidth]{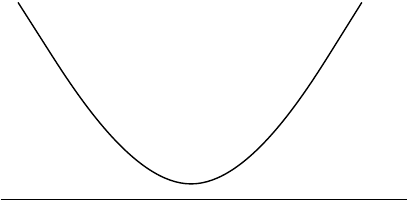}}   
  \quad        
\subfloat[$M=M_*$, one extremal root]{\includegraphics[width=0.3\textwidth]{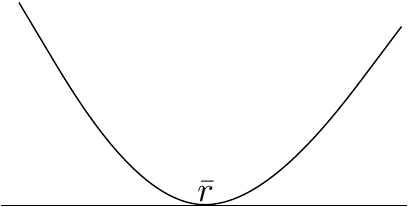}}   
  \quad        
  \subfloat[$M>M_*$, two roots]{\includegraphics[width=0.3\textwidth]{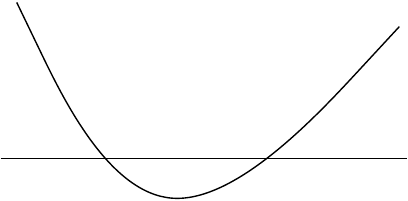}}   
  \quad        
    \caption{The graph of $\Delta_r(r)$ in Kerr-AdS for fixed $a$.}
  \label{fig:TransDeltaRAdS}
\end{figure}

For $M=0$, $\Delta_r$ is always positive and has one extremum at $r=0$ with $\Delta_r(0)=a^2$. The number of local extrema is preserved for all $M$, because otherwise there would be a value of $M$ such that $\Delta_r'=\Delta_r''=0$ has a solution. However, $\Delta_r''$ is positive for all $M$. As we increase $M$, the graph of $\Delta_r$ will deform and cross the horizontal axis for the value of an extremal mass $M=M_*$. Solving $\Delta_r=\Delta_r'=0$, we find that this happens only once, with
\begin{multline}\label{eq:AGivesMass} M_* = \frac{\ell}{\sqrt{6}}\left( - (1+\frac{a^2}{\ell^2})+ \sqrt{ (1+\frac{a^2}{\ell^2})^2+12 \frac{a^2}{\ell^2} }\right)^{1/2} \\ \times
 \left( \frac{2}{3} (1+\frac{a^2}{\ell^2})+ \frac{1}{3}\sqrt{ (1+\frac{a^2}{\ell^2})^2+12 \frac{a^2}{\ell^2} }\right)~.
\end{multline}
The profile of the graph of $\Delta_r$ is shown in figure~\ref{fig:TransDeltaRAdS}. 

We parametrize the values of $(M,a)$ at extremality in terms of the double root $\bar{r}$,
\begin{align}
M_*  &=   \bar{r} \frac{ \left( 1+\frac{\bar{r}^2}{\ell^2}  \right)^2 }{ 1 - \frac{\bar{r}^2}{\ell^2} } \quad, &
a^2_*  &=   \bar{r}^2 \frac{  1+3\frac{\bar{r}^2}{\ell^2} }{ 1 - \frac{\bar{r}^2}{\ell^2} }~.
\end{align}
Note that given $a$ there is always one extremal value for the mass given by \eqref{eq:AGivesMass}. Since the relation $M_*(\bar{r})$ for $0<\bar{r}<\ell $ is one-to-one and onto $\mathbb{R}^+$, there is also a unique critical value of acceleration for any choice of mass. A black hole with shielded singularities requires $M>M_*(a)$ and $r$ is taken larger than the biggest root of $\Delta_r$.

\begin{figure}
  \centering
  \subfloat[$|L|<L_+$, four roots]{\includegraphics[width=0.3\textwidth]{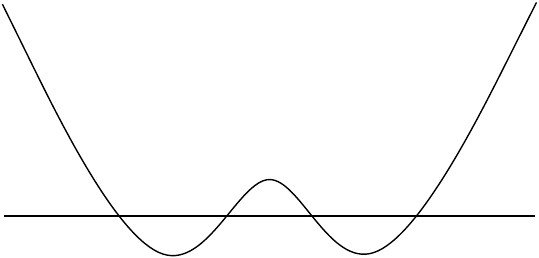}}   
  \quad        
\subfloat[$|L|=L_+$, three roots]{\includegraphics[width=0.3\textwidth]{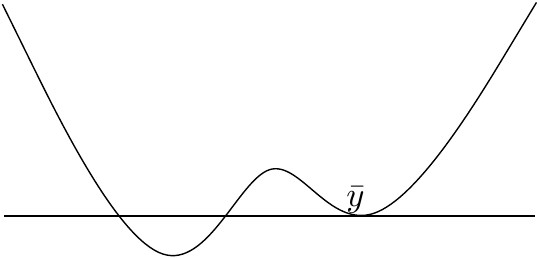}}   
  \quad        
  \subfloat[$|L|>L_+$, two roots]{\includegraphics[width=0.3\textwidth]{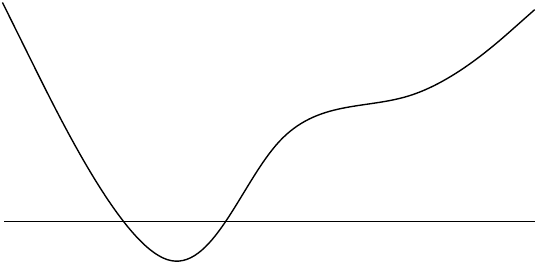}}   
  \quad        
    \caption{Graph of $\Delta_y(y)$ in Kerr-AdS for fixed $a$.}
  \label{fig:TransDeltaLAdS}
\end{figure}

For $L=0$, there are three extrema and four roots to $\Delta_y=0$. We deform the graph by turning on $L$. For some value of $L$ the three extrema will degenerate to one. However, we are interested in the transition from four roots to two roots, which happens at a lower value of $L$. Since $\Delta_y(y=0)=a^2$ for all $L$ and this is the local maximum for $L=0$, the transition from four to two roots involves one of the two local minima, rather than the local maximum, crossing the horizontal line $\Delta_y=0$. 

Solving $\Delta_y=\Delta_y'=0$ we find a unique solution up to sign, $|L|=L_*$ with
\begin{multline} L_* = \frac{\ell}{\sqrt6} \left( 1+\frac{a^2}{\ell^2} + \sqrt{ ( 1+\frac{a^2}{\ell^2} )^2 +12 \frac{a^2}{\ell^2} } \right)^{\frac{1}{2}} \\\times
 \left(\frac{2}{3}( 1+\frac{a^2}{\ell^2}) -\frac{1}{3} \sqrt{ ( 1+\frac{a^2}{\ell^2} )^2 +12 \frac{a^2}{\ell^2} } \right) ~.
\end{multline}
We parametrize the extremal values of $(L,a)$ with respect to the double root $\bar{y}$:
\begin{align}
L_*   &=   \bar{y} \frac{ \left( \frac{\bar{y}^2}{\ell^2} -1 \right)^2 }{ 1 + \frac{\bar{y}^2}{\ell^2} } \quad, &
a^2_* &=   \bar{y}^2 \frac{  3\frac{\bar{y}^2}{\ell^2} -1  }{ 1 + \frac{\bar{y}^2}{\ell^2} }~.
\end{align}
The profile of $\Delta_y$ is shown in figure~\ref{fig:TransDeltaLAdS}.

\subsubsection*{Periodicities}

Let us take $0<|L|<L_*$, in which case there are four roots to $\Delta_y$, $y_1<y_{-}<y_{+}<y_2$. We define $t=t_\pm-y_\pm^2 \psi_\pm$ and $\psi=\psi_\pm$, and expand the metric with $y=y_\pm\mp\rho^2$ close to $\rho^2=0$. At constant $(t_\pm,r)$ the metric becomes
\begin{equation}
  \left. \grad s^2\right|_{r,t_\pm} \approx  \frac{4(r^2+y_\pm^2)}{\mp\Delta_y'(y_\pm)} 
  \times \left(  \grad \rho^2 + \rho^2 \left(\frac{\Delta_y'(y_\pm) }{2}\right)^2  \grad \psi_\pm^2 \right)~.
\end{equation}
Smoothness at $y=y_\pm$ requires the periodicity $\psi_\pm = \psi_\pm+ 2\pi T_\pm $, with
\begin{equation}\label{eq:TpmPeriod} T_\pm =  \frac{2}{| \Delta_y'(y_\pm)|}~. \end{equation}
The two coordinates systems $(t_\pm,\psi_\pm)$ are patched together, away from the roots $y=y_\pm$, by
\begin{subequations}\label{eq:TwoPatchesY1}
\begin{align}
 t_+    & = t_- + (y_+^2-y_-^2) \psi_- ~,\\
 \psi_+ &= \psi_-~.
\end{align} 
\end{subequations}

It follows from \eqref{eq:TpmPeriod} and \eqref{eq:TwoPatchesY1} that at fixed $(y,r)$ the two patches describe torus fibers with periodicities
\begin{equation}\label{eq:NUTLattice} \left(t_\pm,\psi_\pm\right)=\left(t_\pm,\psi_\pm+2\pi\, T_\pm \right) = \left(t_\pm \pm 2\pi\left(y_+^2-y_-^2\right)T_\mp  ,\psi_\pm+ 2\pi\, T_\mp\right) ~.\end{equation} 
We see that a non-zero NUT charge generically necessitates both the existence of closed timelike curves, for instance the curve at fixed $(\psi_\pm,r,y)$, and the non-existence of a global coordinate system with which to describe the $t-\psi$ part of the metric, see~\cite{Misner:1963fr,Moutsopoulos:2009ia}. When $L=0$, \eqref{eq:TwoPatchesY1} becomes $t_+=t_-$ and $\psi_+=\psi_-$, and the periodicity is simply
\begin{equation} \left(t_+,\psi_+\right)=\left(t_+,\psi_++\frac{2\pi}{a\left(1-\frac{a^2}{\ell^2}\right)}\right)~.
\end{equation}
There is however one more case we want to consider, that is when two roots degenerate at $|L|=L_*$.

If there is a double root, say $y_+=y_2=:\bar{y}$, the torus fibers essentially ``uncompactify'' in one direction. This is because expanding close to the double root, the metric is approximately
\begin{multline}
 \left. \grad s^2 \right|_{r,t_\pm} \approx \left[ - \frac{\Delta_r}{r^2+\bar{y}^2} 4\bar{y}^2 + \frac{ \Delta''_y(y_+) }{2} (r^2+\bar{y}^2) \right](y-\bar{y})^2  \grad \psi_+^2  + 2\frac{r^2+\bar{y}^2}{ \Delta''_y(y_+) }\frac{ \grad y^2}{(y-\bar{y})^2} ~.
\end{multline}
The bracket in $g_{\psi_+\psi_+}$ can become negative, but this is inconsequential\footnote{We can equivalently Kaluza-Klein reduce under the isometry generated by $\partial_{t_+}$.}. What is important is that, there is an infinite throat that does not impose any periodicity on $\psi_+$. We still have
\begin{equation}
\left(t_-,\psi_-\right) =\left(t_-,\psi_-+2\pi\, T_-\right)~
\end{equation}
from expanding close to the other root. This enforces the periodicity on the $(t_+,\psi_+)$ patch
\begin{equation}\label{eq:DyExtremalPeriod}
 \left(t_+,\psi_+\right) =\left( t_+ + \left(\bar{y}^2-y_-^2\right) 2\pi\,T_- ,\psi_++ 2\pi\, T_- \right) ~.
\end{equation}
This periodicity is inherited by the polar limit that we define in section \ref{sec:nhek}.

\subsection*{The case of de Sitter}
The Kerr family of black holes in de Sitter space, with cosmological constant $3g^2$, can be obtained from the Kerr anti de-Sitter metric by simply substituting $\ell^2=-1/g^2$. We will consider this case as a side note, with a few more details given in the appendix~\ref{app:dSProfiles}. Here we present the qualitative differences to what was done previously.

We find that unless\footnote{Note that one cannot choose $a^2 \,g^2 \leq 1$ as we did for $\ell^2$, $a^2\leq\ell^2$.
} $a^2 g^2<7 - 2 \sqrt{12}$, the function $\Delta_r$ always has two roots and it is positive in a bounded region between these roots. Furthermore, the ``origin'' $r=0$ is in the bounded region. 
When $a^2 g^2 < 7 - 2 \sqrt{12}$, there are two critical masses $M_\pm$ such that $\Delta_r$ has four roots (three of which are positive) when $M_-(a)<M<M_+(a)$. When $M<M_-$ or $M>M_+$ there are again only two roots and the origin is in between them. The situation for $\Delta_y$ is, in a sense, opposite to that of $\Delta_r$. It always has two roots  unless $a^2 g^2>7 + 2 \sqrt{12}$. For $a^2 g^2>7 + 2 \sqrt{12}$ there are four roots only when $L_-<|L|<L_+$ for two critical values $L_\pm$.

\begin{figure}
\centering
{\includegraphics[width=0.5\textwidth]{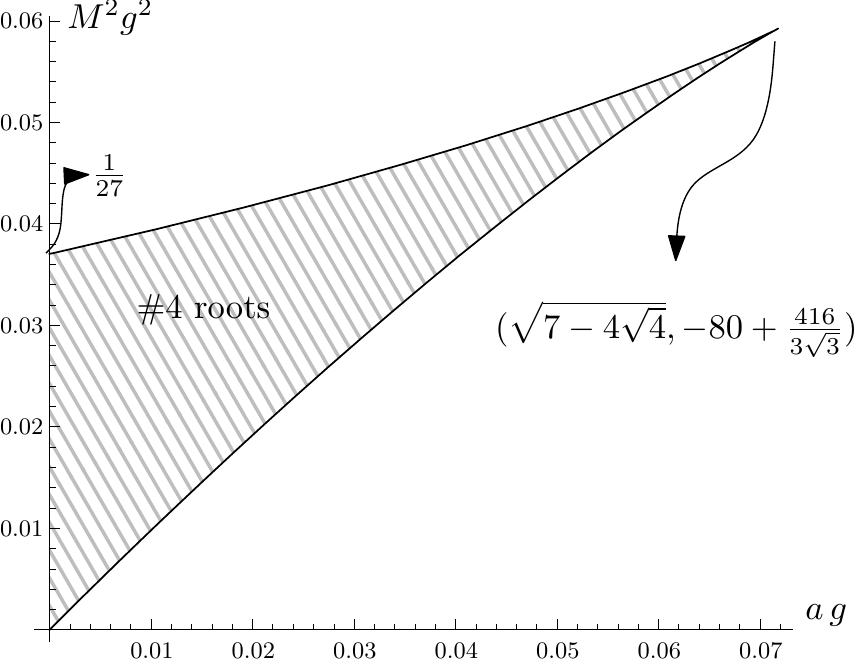}}   
  \quad        
    \caption{Physical regions in the $(M^2,a)$ plane.}
  \label{fig:physMdS}
\end{figure}

A black hole requires that the singularities at $r=0$ are hidden behind a horizon. The parameter space of physical interest is thus given by $a^2g^2<7-4\sqrt{3}$ and $M_-(a)<M<M_+(a)$, as in figure~\ref{fig:physMdS}. The region where $\Delta_r$ is positive covers the exterior of the black hole but within the cosmological horizon~\cite{Gibbons:1977mu}. The periodicity of $(\psi,t)$ is analyzed identically to the negative comoslogical constant case without any surprises. 

\section{Extremal Limits}\label{sec:nhek}
Now that we have introduced the black hole parameters $(a,M,L)$ and how they affect the range of coordinates, we proceed to define the extremal limits. The first limit, which is known as the near-horizon extremal limit, corresponds to blowing up the region close to the extremal horizon, that is, when $M=M_*$. A systematic treatment of the near-horizon limit for supersymmetric backgrounds was presented in~\cite{Reall:2002bh}, but the notion has been known since at least~\cite{Geroch:1969ca}, see also \cite{Racz:1992bp}. We shall also consider a second spacetime limit, which corresponds to blowing up the region of the throat $y=\bar{y}$, when $L=L_*$. For the lack of a better name, we will call this an (extremal) polar limit.

\subsection*{Near-horizon limit}
When the mass attains its lower bound, $M=M_*$, the horizon is extremal in the sense that past and future event horizons do not intersect, equivalently the Hawking temperature is zero. In this case, we expand $\Delta_r$ close to its horizon $r=\bar{r}$ as
\begin{equation}
 \Delta_r  = \frac{1}{\beta^2}(r-\bar{r})^2 + \mathcal{O}(r-\bar{r})^3 
~,\end{equation} where \begin{equation}
 \beta^2 = \frac{1-\frac{\bar{r}^2}{\ell^2}}{1+6\frac{\bar{r}^2}{\ell^2}- 3 \frac{\bar{r}^4}{\ell^2}} ~.
\end{equation}
The metric in the $(t_\pm,\psi_\pm)$ patch is
\begin{multline} \grad s^2 = - \frac{\Delta_r}{r^2+y^2} \left( \grad {t_\pm} + \left(y^2-y_\pm^2\right) \grad \psi_\pm\right)^2
          \\ + \frac{\Delta_y}{r^2+y^2} \left(  \grad {t_\pm} - \left(r^2+ y_\pm^2\right)   \grad \psi_\pm \right)^2
+\frac{r^2+y^2}{\Delta_r}  \grad r^2+\frac{r^2+y^2}{\Delta_y}  \grad y^2~.
\end{multline}
We define the diffeomorphism $(t_\pm,r,\psi_\pm,y)\mapsto(\tau,x,\bar\phi,y)$ for any $\epsilon>0$ by
\begin{subequations} \label{eq:NHEKDiff}\begin{align}
 r&=\epsilon\, x  +\bar{r}~,\\
 t_\pm & = \frac{\beta^2(\bar{r}^2+y_\pm^2)\tau}{\epsilon}~,\\
\psi_\pm &=T_\pm \bar{\phi} + \frac{t_\pm}{(\bar{r}^2+y_\pm^2)}~.
\end{align}\end{subequations}
Taking the limit $\epsilon\rightarrow0^+$ gives the near-horizon extremal Kerr (NHEK) metric
\begin{equation}\label{eq:NHEKmetric}
   \grad \bar{s}^2 = \Omega^2\left( -x^2 \grad \tau^2+\frac{ \grad x^2}{x^2}+\Lambda^2 (T_u  \grad  \bar\phi+x \,\grad \tau)^2\right) + \frac{\bar{r}^2+y^2}{\Delta_y} \grad y^2 ~, 
\end{equation}
where
\begin{align}\label{eq:NHEKParameters}
 \Omega^2 &= \beta^2(\bar{r}^2+y^2) ~,&
 \Omega^2\Lambda^2 &= \frac{\Delta_y}{\bar{r}^2+y^2}4\bar{r}^2\beta^4 ~,&
 T_u &= \frac{\bar{r}^2+y_\pm^2}{2\bar{r}\beta^2} T_\pm~.
\end{align}
We make the observation that if $L\neq0$ then the limit does not inherit a well-defined coordinate range. That is, the lattice in \eqref{eq:NUTLattice} becomes degenerate in the coordinates $(\tau,x,\bar\phi)$ of \eqref{eq:NHEKDiff} when $\epsilon\rightarrow0^+$.  This is reminiscent of the notion of the pinching manifold that was defined in another context in \cite{deBoer:2010ac}. If however $L=0$, then $\bar\phi$ is simply periodic with $\bar\phi=\bar\phi+2\pi$ and the coordinate ranges are well-defined.

The NHEK metric \eqref{eq:NHEKmetric} on a constant slice of $y$ is
\begin{equation}\label{eq:SpacelikeWarpedAdS}
 \left. \grad \bar{s}^2\right|_{y}= \Omega^2\left( -x^2 \grad \tau^2+\frac{ \grad x^2}{x^2}+\Lambda^2 ( \grad  {u}+x\, \grad \tau)^2\right) ~.
\end{equation}
For ${u}\in\mathbb{R}$, this is the so-called spacelike warped AdS$_3$ metric, a metric deformation of AdS$_3$, an exposition of which can be found in~\cite{Jugeau:2010nq}. When $\Lambda^2=1$, the space is precisely AdS$_3$. The coordinates used here are closely related to the self-dual coordinates of~\cite{Coussaert:1994tu}, see appendix~\ref{app:AdS3}. As a group  manifold, $\textrm{AdS}_3=\mathrm{SL}(2,\mathbb{R})$ is preserved by its right-action and left-action, generated respectively by the Killing vectors
\begin{align}
r_0 &=  -\frac{1}{x} \partial_u - \tau\, x \,\partial_x +\frac{1}{2}(1+\frac{1}{x^2}+\tau^2) \partial_\tau\\
r_1 &= x\,\partial_x - \tau \, \partial_\tau \\
r_2 &= \frac{1}{x} \partial_u + \tau \, x \,\partial_x +\frac{1}{2}(1-\frac{1}{x^2}-\tau^2) \partial_\tau \\
\intertext{and}
l_0 &= -\frac{\cosh u}{x}\partial_\tau - x \sinh u \,\partial_x +\cosh u \,\partial_u \\
l_1 &=\frac{\sinh u}{x}\partial_\tau + x\cosh u\, \partial_x -\sinh u \, \partial_u \\
l_2 &= \partial_u~.
\end{align}
However, spacelike warped AdS$_3$, that is when $\Lambda^2\neq1$, is preserved only by the $r_a$, $a=0,1,2$, and $l_2$. The NHEK metric  \eqref{eq:NHEKmetric} is thus also preserved by the Killing vectors $r_a$ and $l_2$. We will henceforth use small latin indices $a,b,c,\ldots$ that take values $0,1,2$ and are raised or lowered with a flat lorentzian metric $\eta_{ab}$. 

Before we move on with the extremal polar limit, let us make a few more remarks. A metric of the form
\begin{equation}\label{eq:NHEKtype}
  \grad \bar{s}^2 = \Omega^2(y) \left( - x^2  \grad \tau^2 +\frac{ \grad x^2}{x^2}+ \Lambda^2(y) ( \grad u+ x \, \grad \tau)^2\right)+ F^2(y)  \grad y^2 ~
 \end{equation}
is precisely AdS$_4$ only if $\Lambda^2=1$. This follows easily from inspection of the curvature. The other two functions, $\Omega^2(y)$ and $F^2(y)$, are uniquely determined up to a diffeomorphism $y\mapsto y'(y)$, e.g. with $F=\ell$ and $\Omega^2=\ell^2\cosh^2 \left(y/\ell^2\right)$. However, the NHEK solution \eqref{eq:NHEKParameters} can have $\Lambda^2=1$ only if $\bar{r}^2=-\ell^2$ and $L=0$. In these coordinates, the AdS$_4$ metric becomes
\begin{equation}
  \grad \bar{s}^2 = \frac{y^2-\ell^2}{4}\left( -x^2 \grad \tau^2+\frac{ \grad x^2}{x^2}+ ( \grad u+x \, \grad \tau)^2\right) + \frac{1}{\ell^2(y^2-\ell^2)} \grad y^2 ~.
\end{equation}
Since the parameter $\bar{r}$ is a positive real number, the NHEK geometry seems to be ``disconnected'' from AdS$_4$. On the other hand, we can ask when does a metric of the form \eqref{eq:NHEKtype} satisfy the Einstein equations of motion. We find that up to diffeomorphisms of $y\mapsto y'(y)$, the most general solution is determined uniquely by two integration constants, see appendix~\ref{app:solvingNHEK}. Therefore, by replacing the parameters $L$ and $\bar{r}^2$ with any real value, the NHEK geometry \eqref{eq:NHEKmetric} with metric functions \eqref{eq:NHEKParameters} is locally the most general Einstein solution of the form \eqref{eq:NHEKtype}.

\subsection*{Polar limit}
Next we consider the extremal limit where we blow up the double root $y=\bar{y}$ of $\Delta_y$ at extremality $L=L_*$. With
\begin{equation}
 \Delta_y  = \frac{1}{\beta^2}(y-\bar{y})^2 + \mathcal{O}(r-\bar{r})^3 
\end{equation} where \begin{equation}
 \beta^2 = \frac{1+\frac{\bar{y}^2}{\ell^2}}{-1+6\frac{\bar{y}^2}{\ell^2}+ 3 \frac{\bar{y}^4}{\ell^2}} ~,
\end{equation}
the metric is
\begin{multline}
 \grad s^2 = - \frac{\Delta_r}{r^2+y^2} \left(  \grad {t_+} + \left(y^2-\bar{y}^2\right)  \grad \psi_+ \right)^2
           + \frac{\Delta_y}{r^2+y^2} \left(  \grad {t_+} - \left(r^2+ \bar{y}^2\right)  \grad \psi_+ \right)^2 \\
+\frac{r^2+y^2}{\Delta_r}  \grad r^2+\frac{r^2+y^2}{\Delta_y}  \grad y^2~.
\end{multline}
We define the diffeomorphism $(t_+, r , \psi_+ ,y)\mapsto(t_+ ,r,\bar\psi,x)$ for any $\epsilon>0$ by
\begin{align}\label{eq:DiffDyLimit}
 \epsilon\, x &= y-\bar{y}\\
\epsilon \, \psi_+ & = \beta^2\bar\psi~.
\end{align}
After taking the limit $\epsilon\rightarrow0^+$ we arrive at the metric
\begin{equation}\label{eq:DyLimitMetric} 
 \grad \bar{s}'^2 = \Omega^2(r)\left( +x^2 \grad \bar\psi^2+\frac{ \grad x^2}{x^2}-\Lambda^2(r) \left(\frac{1}{2\bar{y}\beta^2}  \grad  t_++x \,\grad \bar\psi\right)^2\right) + \frac{\bar{y}^2+r^2}{\Delta_r} \grad r^2 ~,
\end{equation}
with
\begin{align}
 \Omega^2 &= \beta^2(\bar{y}^2+r^2) \\
 \Omega^2\Lambda^2 &= \frac{\Delta_r}{\bar{y}^2+r^2}4\bar{y}^2\beta^4~.
\end{align}
By combining \eqref{eq:DiffDyLimit} amd \eqref{eq:DyExtremalPeriod}, we find that the extremal limit inherits the periodicity
\begin{equation}
 \left(t_+,\bar\psi\right) =\left( t_+ + \right(\bar{y}^2-y_-^2\left)2\pi\,T_-  ,\bar\psi \right)~.
\end{equation}
In this case, the polar extremal limit has a well-defined coordinate range. Recall that this was not the case for the NHEK unless $L=0$.

At constant $r$, the metric in \eqref{eq:DyLimitMetric} becomes
\begin{equation}\label{eq:TimelikeWarpedAdS}
\left. \grad \bar{s}'^2\right|_r=\Omega^2\left( +x^2 \grad \psi^2+\frac{ \grad x^2}{x^2}-\Lambda^2 \left( \grad u+x \,\grad \bar\psi\right)^2\right)~.
\end{equation}
For $u\in\mathbb{R}$ this describes the so-called timelike warped AdS$_3$. For $\Lambda^2=1$ this is again precisely AdS$_3$. However the coordinates of AdS$_3$ in \eqref{eq:TimelikeWarpedAdS} are different to those used for the AdS$_3$ in \eqref{eq:SpacelikeWarpedAdS} for $\Lambda^2=1$. The coordinates of AdS$_3$ used to describe timelike or spacelike ``warping'' should not intimidate the uninitiated reader. We give a brief classification of the various AdS$_3$ coordinates in appendix~\ref{app:AdS3}. Timelike warped AdS$_3$ and by extension the polar limit \eqref{eq:DyLimitMetric} is preserved by only four Killing vectors of AdS$_3$: $r_0$, $r_1$, $r_2$ and $l_0$. Again, the extremal polar limit is locally AdS$_4$ only if $\Lambda^2=1$, which can be obtained from the polar limit with $\bar{r}^2=a^2=\ell^2$ and $M=0$.

\subsection*{Limits for positive cosmological constant}
The extremal limits can also be performed for positive cosmological constant. When $a^2 g^2 < 7-4\sqrt{3}$, the near-horizon limit gives
\begin{equation}\label{eq:NHEKdeSitter}
   \grad \bar{s}^2 = \Omega^2\left( -x^2 \grad \tau^2+\frac{ \grad x^2}{x^2}+\Lambda^2 \left(T_u  \grad  \bar\phi+x \, \grad \tau\right)^2\right) + \frac{\bar{r}^2+y^2}{\Delta_y} \grad y^2~,
\end{equation}
where
\begin{align}
 \Omega^2 &= \beta^2(\bar{r}^2+y^2) ~,&
 \Omega^2\Lambda^2 &= \frac{\Delta_y}{\bar{r}^2+y^2}4\bar{r}^2\beta^4 ~,\\
 T_u &= \frac{\bar{r}^2+y_\pm^2}{2\bar{r}\beta^2} T_\pm~,&
 \beta^2 &= \frac{1+g^2{\bar{r}^2}}{1-6g^2\bar{r}^2- 3 g^4\bar{r}^4} ~.
\end{align}
Here, $\beta^2$ and $\Omega^2$ can now be of either sign, corresponding to the two extremal masses $M=M_\pm$ where the sign of $\Delta_r''(\bar{r})$ is different. Similarly to the case of negative cosmological constant, unless $L=0$, the coordinates are not well-defined by the limit. The extremal polar limit can also be performed in the unphysical region $a^2 g^2 > 7-4\sqrt{3}$ when $|L|=L_\pm$ but we will not persue this here. We simply note that in all cases, the (positive cosmological constant) extremal limits can be obtained from the negative cosmological constant extremal limits with the substitution $g^2=-1/\ell^2$. 

\begin{center}
$\sim$\hspace{-3.7pt}$\sim$\end{center}

We will focus on the negative cosmological constant case. The aim of the subsequent sections is to show that in the near-horizon limit
\begin{equation}
  \grad \bar{s}^2 =  \Omega^2(y)\left( - x^2  \grad \tau^2 +\frac{ \grad x^2}{x^2}+ \Lambda^2(y) \left( \grad u+ x \, \grad \tau\right)^2\right)+ F^2(y)  \grad y^2 ~
\end{equation}
 the space of CKY two-forms is not enhanced from dimension two. Our result can be easily repeated for the polar limit
\begin{equation}
 \grad \bar{s}'^2 =  \Omega^2(r)\left( + x^2  \grad \psi^2 +\frac{ \grad x^2}{x^2} - \Lambda^2(r)\left( \grad u+ x \, \grad \psi\right)^2\right)+ F^2(r)  \grad r^2 ~. 
\end{equation}

\section{Holonomy of extremal limits}\label{sec:noparallel}
A parallel two-form, $\nabla_\mu  K_{\nu\rho}=0$, solves trivially the conformal Killing-Yano equation. Although, the black hole metrics do not allow parallel \newadd{$p$-forms}, 
the situation could potentially change for the - highly symmetric - extremal limits. In this section we cast away with such doubt and show that this is not the case. We will work primarily with the near-horizon geometry, but the result also applies  to the polar extremal limit.

If there is a parallel two-form $K$, then one has the integrability condition
\begin{equation}\label{eq:CurvOnTwo}
[ \nabla_\mu,\nabla_\nu] K_{\rho\sigma} = R_{\mu\nu\rho}{}^\tau K_{\sigma\tau} +  R_{\mu\nu\sigma}{}^\tau K_{\tau\rho} = 0 ~.
\end{equation}
That is, parallel two-forms are stabilized by the curvature at any point, and more generally by the holonomy algebra of the Levi-Civita connection. Our main task is to show that the holonomy algebra, which to first order is generated by the curvature, is not special but spans $\mathfrak{so}(1,3)$. Since the curvature acts on two-forms in the adjoint representation and  $\mathfrak{so}(1,3)$ has no central elements, there can be no parallel two-forms.

Let us first introduce the set of one-forms:
\begin{align}
\theta^0 &= -x \cosh u \,  \grad \tau + \frac{\sinh u}{x}  \grad x \\
\theta^1 &= \frac{\cosh u}{x}  \grad x - x \sinh u \, \grad \tau\\
\theta^2 &=  \grad u+x \, \grad \tau ~.
\end{align}
A relation we shall use soon is the Maurer-Cartan structure equation
\begin{equation}\label{eq:dthetaMC} \grad \theta^a -\frac{1}{2}\epsilon^a{}_{bc}\theta^b\wedge\theta^c=0 ~.\end{equation} 
The $\theta^a$ form a basis for the right-invariant Maurer-Cartan one-forms of $\mathrm{SL}(2,\mathbb{R})$. In particular, the metric of spacelike warped AdS$_3$ is equal to
\begin{equation}\label{eq:AdSMetricWithTildes}
 ds^2_{\textrm{wAdS}_3} =  \Omega^2\left( - \theta^0 \otimes \theta^0 + \theta^1 \otimes \theta^1 +\Lambda^2 \theta^2 \otimes \theta^2 \right)~. \end{equation}
The $\theta^a$ are dual to the right-invariant Killing vectors of AdS$_3$ by $\theta^a(l_b)=\delta^a_b$. Notice how, the right-action of $r_a$ leaves the basis $\theta^a$ invariant, whereas $l_2$ preserves separately the first two and the last one combination in the summand of the spacelike warped AdS$_3$ metric. When $\Lambda^2=1$, the metric of AdS$_3$ can also be written similarly to \eqref{eq:AdSMetricWithTildes} in terms of the left-invariant Maurer-Cartan one-forms $\tilde\theta^a$, for which $\tilde\theta^a(r_b)=\delta^a_b$. This implies that $\tilde\theta^a=\tilde{M}^a{}_b\theta^b$ for a matrix $\tilde{M}_a{}^b$ that is an element of $\mathrm{O}(1,2)$. Since $\tilde{M}_{ac}\tilde{M}^c{}_b=\eta_{ab}$ we have also $\tilde{M}_{ab}=\theta_b(r_a)$.

It is convenient to define an orthonormal basis $\hat\theta^A$ for a metric of the form 
\begin{equation}\label{eq:NHEKmetricExp}
  \grad \bar{s}^2= e^{2\omega(y)} \left( - x^2  \grad \tau^2 +\frac{ \grad x^2}{x^2}+ e^{2\lambda(y)} ( \grad u+ x \, \grad \tau)^2\right)+ e^{2f(y)}  \grad y^2 ~
 \end{equation}
by using the one-forms of AdS$_3$:
\begin{equation}\label{eq:NHEKOrthoFrame}\begin{aligned} 
\hat\theta^0&=e^{\omega}\theta^0 ~,& 
\hat\theta^1 &=e^{\omega}\theta^1 ~,&
\hat\theta^2&=e^{\omega+\lambda}\theta^2~,& 
\hat\theta^y &=e^{f} dy~. 
\end{aligned}\end{equation}
For future use we collect the first three relations in $\hat\theta^a=\hat{M}^a{}_b \theta^b$ by defining the diagonal matrix $\hat{M}^a{}_b$. We first calculate the spin coefficients from
\begin{equation}\label{eq:justaspin}
\grad \hat\theta^A + \hat\omega^A{}_B \wedge \hat\theta^B =0~
\end{equation}
and then calculate the curvature.

By using \eqref{eq:dthetaMC}, the solution to \eqref{eq:justaspin} is found to be
\begin{equation}\label{eq:SpinConnection}
 \begin{aligned}
\hat\omega^0{}_1 &= (- e^{-\omega-\lambda}+\frac{1}{2} e^{-\omega+\lambda})\hat\theta^2 &
\hat\omega^2{}_0 &= \frac{1}{2} e^{-\omega+\lambda}\hat\theta^1 \\
\hat\omega^1{}_2 &= \frac{1}{2} e^{-\omega+\lambda} \hat\theta^0 &
\hat\omega^0{}_z &= \dot\omega e^{-f}\hat\theta^0\\
\hat\omega^1{}_z &= \dot\omega e^{-f}\hat\theta^1 &
\hat\omega^2{}_z &= (\dot\omega+\dot\lambda)e^{-f}\hat\theta^2~.
\end{aligned}
\end{equation}
The curvature two-form $R^\nabla{}_{AB}$ is calculated from the right-hand side of
\[ R^\nabla{}_{AB}= \frac{1}{2} R_{ABMN}\,  \grad x^M\wedge  \grad x^N =  \grad \hat\omega_{AB}+\hat\omega_{AC}\wedge \hat\omega^C{}_B ~. \]
We find the following components
\begin{subequations}\label{eq:CurvTwoNhek}\begin{align}
R^\nabla{}^1{}_0 &= \dot\lambda e^{-\omega+\lambda-f} \hat\theta^y \wedge \hat\theta_2 
- \left( e^{-2\omega}-\frac{3}{4} e^{-2\omega+2\lambda}+\dot\omega^2 e^{-2f} \right)\hat\theta^1\wedge\hat\theta_0 ~,\\
R^\nabla{}^2{}_y &= {\dot\lambda} e^{-\omega+\lambda-f} \hat\theta^0\wedge \hat\theta^1 
+ \left(\ddot\omega+\ddot\lambda+(\dot\omega+\dot\lambda)(\dot\omega+\dot\lambda-\dot{f})\right)e^{-2f} \hat\theta^y\wedge \hat\theta^2 ~,\\
R^\nabla{}^0{}_y &=  \frac{\dot\lambda}{2} e^{-\omega+\lambda-f}\hat\theta^1\wedge \hat\theta^2 + \left( \ddot\omega +\dot\omega^2-\dot\omega\dot{f} \right) e^{-2 f} \hat\theta^y\wedge \hat\theta^0 ~,\\
R^\nabla{}^2{}_1 &=-\frac{\dot\lambda}{2} e^{-\omega+\lambda-f}\hat\theta^y\wedge \hat\theta^0 
+ \left( \dot\omega(\dot\omega+\dot\lambda)e^{-2f}+\frac{1}{4} e^{-2\omega+2\lambda} \right) \hat\theta^1\wedge \hat\theta_2 ~,\\
R^\nabla{}^2{}_0 &= \frac{\dot\lambda}{2} e^{-\omega+\lambda-f}\hat\theta^y\wedge \hat\theta^1 
-\left( \frac{1}{4} e^{-2\omega+2\lambda}+\dot\omega(\dot\omega+\dot\lambda) e^{-2f}\right)\hat\theta^0\wedge\hat\theta_2 ,\\
R^\nabla{}^1{}_y &=\frac{\dot\lambda}{2} e^{-\omega+\lambda-f}\hat\theta^0\wedge \hat\theta_2 
+ \left(\ddot\omega+\dot\omega(\dot\omega-\dot{f})\right) e^{-2f} \hat\theta^y\wedge \hat\theta^1 ~.
\end{align}\end{subequations}
Notice that the metric is locally AdS$_4$, $R^\nabla{}_{AB}=-\frac{1}{4\ell^2}\hat\theta^A\wedge\hat\theta^B$, only if $\dot{\lambda}=0$. 

The curvature two-form acts on two-forms linearly through the adjoint action,
\[ \hat\theta^A\wedge\hat\theta^B : \hat\theta^C \wedge \hat\theta^D \longmapsto \eta^{BC}\hat\theta^A\wedge\hat\theta^D  -\eta^{BD}\hat\theta^A\wedge\hat\theta^C
-\eta^{AC}\hat\theta^B\wedge\hat\theta^D+ \eta^{AD}\hat\theta^B\wedge\hat\theta^C~,\]
as is consistent with the integrability condition \eqref{eq:CurvOnTwo}. From the form of \eqref{eq:CurvTwoNhek}, the span of $R^\nabla_{cy}$ and $R^\nabla_{ab}$, where $a,b,c$ are all different, is either the direct sum of $\langle\hat\theta^a\wedge\hat\theta^b\rangle$ and $\langle\hat\theta^a\wedge\hat\theta^y\rangle$ or a one-dimensional subspace thereof, depending on the determinant of the matrix transformation in
\begin{equation}
\begin{pmatrix}
R^\nabla_{cy} \\ R^\nabla_{ab}
\end{pmatrix}=
\begin{pmatrix}
R_{cyab} & R_{cycy} \\
R_{abab} & R_{abcy}
\end{pmatrix}
\begin{pmatrix}
 \hat\theta^a\wedge\hat\theta^b \\ \hat\theta^a\wedge\hat\theta^y
\end{pmatrix}
\end{equation}
All together, the $R^\nabla_{AB}$ span $\Lambda^2=\mathfrak{so}(1,3)$ if and only if
\begin{equation}\label{eq:CurvDet}
 R_{abab}R_{cycy}-(R_{abcy})^2 \neq 0
\end{equation}
for all permutations of $(a,b,c)=(0,1,2)$. 
So far we have not used the specific functions $(\lambda,\omega,f)$ of the NHEK geometry.  We can check \eqref{eq:CurvDet} at any point using the NHEK solution \eqref{eq:NHEKParameters} and a computer calculation confirms its validity. Since $\mathfrak{so}(1,3)$ is centreless, there can be no two-form that is stabilized by the curvature two-form.  Whence, there are no parallel two-forms in the NHEK geometry. This result applies only to the NHEK solution, that is the most general Einstein solution of this form.

The derivation can be repeated for the polar limit, which is of the form
\begin{equation}\label{eq:POLARmetricExp}
  \grad \bar{s}'^2 =  e^{2\omega(y)} \left( x^2  \grad \psi^2 +\frac{ \grad x^2}{x^2} - e^{2\lambda(y)} ( \grad u+ x \, \grad \psi)^2\right)+ e^{2f(y)}  \grad y^2 ~.
 \end{equation}
As we commented earlier, the geometry at constant $y$ is the so-called timelike warped AdS$_3$, so we can similarly use the right-invariant one-forms $\theta^a$ adapted to the timelike warping. The equivalent orthonormal frame to \eqref{eq:NHEKOrthoFrame} is now
\begin{align} 
\hat\theta^0&=e^{\omega+\lambda}\left(  \grad u + x \, \grad \psi\right) ~,& 
\hat\theta^1 &=e^{\omega}\left( \cos u\, x \, \grad \psi + \sin u \frac{ \grad x}{x}\right) ~,\\
\hat\theta^2&=e^{\omega}\left(-\sin{u}\, x\,  \grad \psi + \cos u \frac{ \grad x}{x}\right)~,& 
\hat\theta^y &=e^{f}  \grad y~. 
\end{align}
We simply remark here that the curvature two-form components are the same as in \eqref{eq:CurvTwoNhek} with the interchange of flat indices $2\leftrightarrow 0$. This explains why we chose, perhaps mysteriously, to present the curvature two-form in \eqref{eq:CurvTwoNhek} with some flat indices up and others down. The algebraic relation \eqref{eq:CurvDet} can then be used in place, giving the same result that there are no parallel two-forms in the extremal limit.

\section{Larger than two is eight}\label{sec:LargerIsEight}
In \cite{Rasmussen:2010rw}, Rasmussen showed that the Killing-Yano two-form of the Kerr-(A)dS black hole, including NUT charge, survives the near-horizon limit. Its Hodge dual is the so-called principal CKY two-form of the geometry $K_{\textit{p}}$. That is, it is given by $K_{\textit{p}}=\grad b$, where
\begin{equation}
 b = -\frac{y^2+\bar{r}^2}{2} x \,\grad t - \frac{y^2}{2}\, \grad u~.
\end{equation}
A similar result holds for the extreme polar limit: there are two CKY two-forms, $K_{\textit{p}}=\grad b$ and $\ast K_{\textit{p}}$, where
\begin{equation}
 b = -\frac{r^2+\bar{y}^2}{2} x\,\grad\bar\psi - \frac{r^2}{2}\, \grad u~.
\end{equation}
For definiteness we will work with the near-horizon limit, but the section can be read for the polar limit instead.

In this section we shall prove that there are either only two linearly independent CKY two-forms, $K_\textit{p}$ and $\ast K_\textit{p}$, or else the space of CKY two-forms $\mathcal{K}$ is 8-dimensional. In the latter case the rank of $\mathcal{K}$ is the maximal allowed, that is twice the number of independent Killing vectors, the $r_a$ and $\partial_u$. Our ultimate aim is to show that the maximal case, $\dim\mathcal{K}=8$, is not realized.

Recall that all CKY two-forms are mapped to a pair of Killing vectors. An explicit calculation shows that
\begin{equation}
 \ast  \grad \ast  \grad  b = 3\,T_u^2\, g(\partial_u,-)~,
\end{equation}
and so we find that the known CKY two-forms are such that $\kto{\ast K_\textit{p}}{T_u^2 \partial_u }{0}$ and $\kto{K_\textit{p}}{0}{-T_u^2 \partial_u }$. Since the space of CKY two-forms $\mathcal{K}$ is a vector space, either these two CKY two-forms span the entire space or the space is bigger. If $\dim\mathcal{K}>2$ then there is at least one CKY two-form $K$ such that
$\kto{K}{\tilde{r}}{r}$ with
\begin{align}
 \tilde{r} &= \tilde{A} \, r_0 + \tilde{B} \, r_1 + \tilde{C} \, r_2 ~, \label{eq:TildeR}\\
 r &= A \, r_0 + B \, r_1 + C \, r_2 ~.\label{eq:R}
\end{align}
The $r_a$ transform in the adjoint (vector) representation $V$ of $\mathfrak{so}(1,2)$. It will then follow that the image $\pi\left(\mathcal{K}/\langle K_\textit{p},\ast K_\textit{p}\rangle\right)$ is the entire $V\oplus V$. 

Assume a CKY two-form $K$ as before, $\kto{K}{\tilde{r}}{r}$. If $r$ and $\tilde{r}$ are linearly dependent, $r=c\,\tilde{r}$ with $c\neq0$, then we also have the CKY two-form
$$K'=\kto { \frac{1}{c^2+1}( K - c \ast K  )}{\tilde{r}}{0}~.$$
If on the other hand $r$ and $\tilde{r}$ are non-zero and linearly independent, the CKY two-form $K'=\lie_r K$ is such that $\kto{K'}{r'}{0}$ with $r'=[r,\tilde{r}]$ non-zero\footnote{
It follows from the properties of $\mathfrak{so}(1,2)$ that $r'$ cannot be zero, $\mathrm{stab}(r)=\langle r \rangle$.
}. Finally, if either $r$ or $\tilde{r}$ are zero, the action of Hodge duality allows us to consider the case where $\kto{K}{\tilde{r}}{0}$ in any case. The vector representation $V$ is irreducible, though. Whence, through the action of $\mathfrak{so}(1,2)$ and linearity, all $\tilde{r}\in V$ can be obtained.

We can be more explicit by using the properties of $V$. The action of $\mathfrak{so}(1,2)$ can be integrated on both sides of $\kto{\lie_\xi K}{\lie_\xi\tilde{r}}{0}$. Then any $\tilde{r}$ as in \eqref{eq:TildeR} can be transformed into a Killing vector $\tilde{r}'$ proportional to $r_0$, $r_2$ or $r_0\pm r_2$, depending on whether the length $\tilde{A}^2-\tilde{B}^2-\tilde{C}^2$ is respectively\footnote{It is useful to think of the familiar Lorentz transformations acting on the three-dimensional Minkowski space $\langle r_a \rangle = \mathbb{R}^{1,2}$, with $\mathfrak{sl}(2,\mathbb{R})=\mathfrak{so}(1,2)$.} positive, negative, or zero. We can then act on $\kto{K'}{\tilde{r}'}{0}$ with any of the $r_a$ so that we obtain all of the characteristic elements of $V$ under the action of $\mathfrak{so}(1,2)$. By the action of Hodge duality, the same is true for the closed conformal Killing-Yano two-forms.

To summarize, either there are two CKY two-forms as found in \cite{Rasmussen:2010rw}, or the space of CKY two-forms $\mathcal{K}$ is augmented so that its image under the map $\pi$ spans the entire double copy of the space of Killing vectors $V\oplus\mathbb{R}$. In the latter case, we showed that the Killing-Yano two-forms transform in the same representation as $ V\oplus\mathbb{R}$, as do the the closed conformal Killing-Yano two-forms. Explicitly, if there are more than two independent CKY two-forms, then there are three independent Killing-Yano two-forms $K_a$ with $\kto{K_a}{r_a}{0}$ for each $a=0,1,2$. This will allow us to write down an Ansatz for the most general $K\in\mathcal{K}/\langle K_\textit{p},\ast K_\textit{p}\rangle$ that fails to satisfy the CKY equation.

Part of these results can be immediately generalized. Given a four-dimensional Einstein manifold $(M^4,g)$ and a reductive isometry algebra $\mathfrak{g}= \oplus_i^N \mathfrak{g}_i$, that is each $\mathfrak{g}_i$ is a simple Lie algebra, the space of conformal Killing-Yano two-forms modulo the space of parallel two-forms can be decomposed under $\mathfrak{g}$ into a direct sum of a subset of the prime ideals $\oplus_{i\in S}\mathfrak{g}^\mathcal{K}_i$, where each $\mathfrak{g}^\mathcal{K}_i$ is either $\mathfrak{g}_i\oplus\mathfrak{g}_i$ or $\mathfrak{g}_i$. The first case, $\mathfrak{g}^\mathcal{K}_i= \mathfrak{g}_i \oplus \mathfrak{g}_i$, is when the CKY two-forms can be decomposed into Killing-Yano and closed conformal Killing-Yano two-forms. The second case, $\mathfrak{g}^\mathcal{K}_i= \mathfrak{g}_i$, is when the two-forms cannot be decomposed like that, while Goursat's lemma associates to each such case  an automorphism $\mathfrak{g}_i\rightarrow \mathfrak{g}_i$. However, we cannot always decompose the CKY two-forms into Killing-Yano forms and closed conformal Killing-Yano forms as we did for the extremal limits.

\section{No more CKY two-forms}\label{sec:Ansatz}
In this section we work with the assumption that there is a  closed conformal Killing-Yano two-form $K_a$ such that 
\begin{equation}\label{eq:KaMap}
\kto{K_a}{0}{r_a}~. 
\end{equation}
The CKY two-form equation \eqref{eq:CKY2eqInd} becomes
\begin{equation}
\nabla_{X}K_a  = X^\flat \wedge \left(r_a\right)^\flat.\label{eq:CKYeq-1}
\end{equation}
We first work with \eqref{eq:KaMap} in order to derive an Ansatz for $K_a$ that fails to pass the test of \eqref{eq:CKYeq-1}. This is an Ansatz in the sense that, if $K_a$ satisfies \eqref{eq:KaMap} then it has to be of this form. The failure of the Ansatz to pass the test proves that there are only two CKY two-forms, $K_\textit{p}$ and its Hodge dual.

\subsection*{Birth of an Ansatz}

The $n$-th action of the Lie derivative on $K_a$ along a Killing vector $r$, where $r$ is a linear combination of the right-acion as in \eqref{eq:R}, is also a CKY two-form with $\kto{\lie_r^n K_a}{0}{\lie_r^n {r}_a}$. If we integrate this we arrive at
\begin{equation}\label{eq:KaIntegMap}
 \kto{e^{\epsilon \lie_r}K}{0}{e^{\epsilon \lie_r}{r}_a}~.
\end{equation}
We can exponentiate the adjoint action $[r_a,r_b]=-\epsilon_{ab}{}^c r_c$ and define the matrix $S_\epsilon{}_a{}^b\in\mathrm{SO}(1,2)$ by $e^{\epsilon \lie_r}{r}_a= S_\epsilon{}_a{}^b r_b$. Subtracting \eqref{eq:KaIntegMap} from $S_\epsilon{}_a{}^b$ times \eqref{eq:KaMap} we get
\begin{equation}
\kto{  e^{\epsilon\lie_r} K_a - S_\epsilon{}_a{}^b K_b }{ 0 }{ 0}~.
\end{equation}
We have shown that there are no parallel two-forms and so
\begin{equation}\label{eq:KaIsSK}
 e^{\epsilon\lie_r} K_a = S_\epsilon{}_a{}^b K_b~.
\end{equation}

In order to proceed, we make the observation that the $r_a$ act transitively at constant $y$ on the near-horizon geometry. The orbit is the space known as spacelike warped AdS$_3$, which is diffeomorphic (as a manifold) to $\mathrm{SL}(2,\mathbb{R})$. The infinitesimal action on a group can be integrated for all $\epsilon\in\mathbb{R}$, which defines the flow $\phi_\epsilon:w\textrm{AdS}_3\rightarrow w\textrm{AdS}_3$ according to $r(f)|_p=\left.\frac{\grad}{\grad t} \left( f\circ\phi_\epsilon\right)\right|_{t=0}$ for any function $f$ and at any point $p$, see e.g. \cite{Isham:1999rh}. On a vector field $X$ and a one-form $a$, the exponential of the Lie derivative is related to the push-forward and pullback of $\phi_\epsilon$, respectively, by
\begin{align}
 e^{\epsilon \lie_r}  \left.X\right|_p & = \phi_{-\epsilon}{}_\ast \left.X\right|_{\phi_{\epsilon}(p)} \\
 e^{\epsilon \lie_r}\left. a\right|_p & = \phi_{\epsilon}{}^\ast\left. a\right|_{\phi_{\epsilon}(p)} ~.
\end{align}
We henceforth fix a slice $y=y_0$ and a point $p\in w\textrm{AdS}_3$. At any other point $\phi_\epsilon(p)\in w\textrm{AdS}_3$ of the same slice of $y$, \eqref{eq:KaIsSK} implies that $K_a$ is given by
\begin{equation}\label{eq:KaIsPullBack}
 \left.K_a\right|_{\phi_\epsilon(p)} = S_{\epsilon}{}_a{}^b \phi_{-\epsilon}{}^\ast\left. K_b\right|_p~.
\end{equation}
This equation is what allows us to write an Anzatz for $K_a$.

Next, we need an expression for the transformation matrix $S_\epsilon{}_a{}^b$. The $r_a$ transform the same as their dual one-forms, $\tilde\theta^a(r_b)=\delta^a_b$, whereas according to the discussion above \eqref{eq:AdSMetricWithTildes}, we have the relation $\tilde\theta^a = \tilde{M}^a{}_b \theta^b$ where $\tilde{M}_{ab}$ is an $\mathrm{O}(1,2)$ matrix and the $\theta^a$ are invariant under the action of the $r_a$. By contracting the relation with $r_c$ we arrive at $\tilde{M}_{ab}=\theta_b(r_a)$. Now consider the relation $\left. \tilde\theta^a \right|_{p} = \tilde{M}_p {}^a{}_b  \left. \theta^b \right|_{p}~$ at a fixed point $p$. It transforms as $
 S_\epsilon{}_a{}^b  \left. \tilde\theta_b \right|_{p} =
 \tilde{M}_{\phi_\epsilon(p)} {}^a{}_b  \left. \theta^b \right|_{p}~$. We hence deduce that
\begin{equation}\label{eq:SIsM}
 S_\epsilon{}_a{}^b = \tilde{M}_{\phi_\epsilon(p)}{}_a{}^c \tilde{M}_{p}{}^b{}_c = \left.\theta^c(r_a)\right|_{\phi_\epsilon(p)} \tilde{M}_p{}^b{}_c ~,
\end{equation}
which we can insert in \eqref{eq:KaIsPullBack}.

At fixed $y$ and fixed $p\in w\textrm{AdS}_3$ we write $K_a$ as 
\begin{equation}\label{eq:KaIs}
 \left.K_a \right|_p= H(y)_{ab}  \frac{1}{2} \epsilon^b{}_{cd} \left.\hat\theta^c \right|_p\wedge \left.\hat\theta^d \right|_p+ G(y)_{ab}\left. \hat\theta^y \right|_p\wedge\left. \hat\theta^b \right|_p~.
\end{equation}
At any other point $\phi_\epsilon(p)$ of the same slice of $y$, $K_a$ is given by \eqref{eq:KaIsPullBack}:
\begin{multline}\label{eq:PreAnsatz}
 \left. K_a \right|_{\phi_\epsilon(p)} = S_\epsilon{}_a{}^b \Big(  H(y)_{bc} \frac{1}{2} \epsilon^c{}_{de}\, \phi_{-\epsilon}{}^\ast \left.\hat\theta^d \right|_p \wedge  \phi_{-\epsilon}{}^\ast \left. \hat\theta^e \right|_p \\+ 
G(y)_{bc} \, \phi_{-\epsilon}{}^\ast \left. \hat\theta^y \right|_p \wedge \phi_{-\epsilon}{}^\ast  \left.\hat\theta^c \right|_p \Big)
~.
\end{multline}
However the orthonormal basis $\hat\theta^A$ is given by \eqref{eq:NHEKOrthoFrame} and in particular it is right-invariant. If we also use the expression for $S_\epsilon{}_a{}^b$ in \eqref{eq:SIsM} and absorb the matrix $\tilde{M}_p{}^a{}_b$ multiplying the left of $H(y)_{ab}$ and $G(y)_{ab}$ into their definition, then \eqref{eq:PreAnsatz} becomes
\begin{equation}\label{eq:KaAnsatz}
K_a = \theta^b(r_a) \left(  H(y)_{bc} \frac{1}{2} \epsilon^c{}_{de} \hat\theta^d  \wedge \hat\theta^e + 
G(y)_{bc} \hat\theta^y \wedge \hat\theta^c \right)~,
~
\end{equation}
which is valid at any point of the fixed slice $y$. Furthermore, the equality should vary smoothly over $y$. Our Ansatz is thus that $K_a$ is given by \eqref{eq:KaAnsatz}, with unknown functions of $y$ the matrices $H(y)_{ab}$ and $G(y)_{ab}$. It should hold for any CKY two-form that satisfies our initial assumption \eqref{eq:KaMap}. 

\subsection*{Death of the Ansatz}
We have come a long way since we wrote down the CKY defining equation. So far we have shown that, there are either two independent CKY two-forms in the NHEK background, or else there are three more linearly independent Killing-Yano two-forms of the form \eqref{eq:KaAnsatz}. It is a straightforward calculation to check whether \eqref{eq:KaAnsatz} satisfies the CKY equation. We will explicitly demonstrate here that, unless a background of the form \eqref{eq:NHEKmetricExp} is precisely (locally) AdS$_4$, there is no solution to the matrices $H(y)_{bc}$ and $G(y)_{bc}$.

We begin with taking the derivative of $K_a$ with respect to $X=\partial_y$, in which case the right-hand side of the CKY equation \eqref{eq:CKYeq-1} becomes
\begin{equation}\label{eq:CKY-eqz1}
\nabla_{y}K_a =  \hat\theta_b(r_a) e^f \hat\theta^y\wedge \hat\theta^b~.
\end{equation}
Let us use the diagonal matrix $\hat{M}$ by $\hat\theta^a=\hat{M}^a{}_b \theta^b$, that is from \eqref{eq:NHEKOrthoFrame}
\begin{equation}
\hat{M}^a{}_b = e^\omega \begin{pmatrix} 1 &0 &0 \\ 0& 1 &0 \\ 0&0& e^\lambda \end{pmatrix}~.\end{equation}
We observe that the spin connection has no $y$-component, $\omega_{AB}(\partial_y)=0$, so $\nabla_y$ acts in the orthonormal basis as $\partial_y$. By using the Ansatz \eqref{eq:KaAnsatz} and comparing to \eqref{eq:CKY-eqz1} we find
\begin{subequations}\label{eq:HandG1}\begin{align}
\partial_y H_{bc} & = 0\\
\partial_y G_{bc} &=  e^f \hat{M}_{cb}~.
\end{align}\end{subequations}
At this point, $H_{bc}$ is a constant matrix and the $y$-dependence of $G_{bc}$ is fixed.

Next, we take the covariant derivative of $K_a$ with respect to $l_b$. We can use the spin connection as found in \eqref{eq:SpinConnection} and $\theta^a(l_b)=\delta^a_b$. The right-hand side of the CKY equation is
\begin{equation}\label{eq:CKY-eqz2}
\nabla_{l_b} K_a =  (l_b)^\flat\wedge(r_a)^\flat = \theta^e(r_a) \hat{M}_{cb}\hat{M}_{de} \, \hat\theta^c\wedge \hat\theta^d~. \end{equation}
In order to calculate the derivative $\nabla_{l_b} K_a$ we make use\footnote{This follows from $\grad f(Y)=\lie_Y f$ with $Y=l_a$ and $f=\tilde{M}_{ab}=\theta_b(r_a)$.} of $\grad \tilde{M}_{ab}=\grad \left(\theta_b(r_a)\right)=\tilde{M}_{ad}\epsilon^{d}{}_{bc}\theta^c$. We thus calculate the left-hand side of \eqref{eq:CKY-eqz2} as
\begin{equation}\label{eq:nablalLHS}
\begin{aligned}
\nabla_{l_b}K_a &=
\tilde{M}_a{}^c \Big( \epsilon_{cdb} \frac{1}{2} H^d{}_e \epsilon^e{}_{fg} 
- H_{cd} \epsilon^d{}_{eg} \hat\omega^e{}_f(l_b)
- G_{cg} \hat\omega^y{}_f(l_b)\Big)\hat\theta^f\wedge\hat\theta^g\\
&+\tilde{M}_a{}^c \Big( \epsilon_{cdb} G^d{}_e 
- H_{cd} \epsilon^d{}_{fe} \hat\omega^{f}{}_y(l_b) 
- G_{cd} \hat\omega^d{}_e(l_b)  \Big)\hat\theta^y\wedge \hat\theta^e 
\end{aligned}
\end{equation}
Equating \eqref{eq:CKY-eqz2} and \eqref{eq:nablalLHS} gives two sets of equations\footnote{We use $\epsilon^{abc}\epsilon_{abd}=-2\delta^c_d$ and $\epsilon^{abc}\epsilon_{a'b'c}=2\delta^{[a}_{a'} \delta^{b]}_{b'}$}
\begin{align}
\epsilon_{abd}H^d{}_c - H_{ad}\hat\omega_c{}^d(l_b) - G_{ad}\hat\omega^y{}_e(l_b) \epsilon^{ed}{}_c &= \hat{M}_{ad}\hat{M}_{eb} \epsilon^{ed}{}_c \label{eq:nablalk1}~,\\
\epsilon_{adc}G^d{}_b - H_{ad}{}\epsilon^d{}_{eb}\hat\omega^e{}_y(l_c) - G_{ad}\hat\omega^d{}_b(l_c) &=0 \label{eq:nablalk2}
\end{align}

Setting $b=c$ in these two equations, we find that the off-diagonal components of $G_{ab}$ and $H_{ab}$ are zero. It is easy to see this. The elements $\hat\omega^e{}_y(l_b)$ are non-zero only when $e=b$ and the $\hat\omega^e{}_d(l_b)$ are non-zero only when the $e,d,b$ are all different. So with $b=c$ only the first terms, $\epsilon_{adc}G^d{}_b$ and $\epsilon_{abd}H^d{}_c$, survive and they give that the off-diagonal $G_{ab}$ and $H_{ab}$ are zero. The same result is obtained whenever one or more of the $a,b,c$ are the same. The rest of the equation components are
\begin{subequations}\label{eq:GHall}
\begin{align}
H_{22}+ H_{00}\frac{1}{2} e^\lambda + G_{00} \dot\omega e^{\omega-f} & =-e^{2\omega}~,\\
-H_{11} - H_{00} \left( 1 -\frac{1}{2}e^{2\lambda} \right) - G_{00} (\dot\omega+\dot\lambda) e^{-f+\omega+\lambda} & = e^{2\omega+\lambda} ~,\\
-H_{22} + H_{11}\frac{1}{2} e^\lambda + G_{11}\dot\omega e^{-f+\omega} &= e^{2\omega}~, \\
-H_{00} - H_{11} \left( 1 - \frac{1}{2} e^{2\lambda} \right) - G_{11} (\dot\omega+\dot\lambda)e^{-f+\omega+\lambda} &= -e^{2\omega+\lambda} \label{eq:GHone} ~,\\
H_{11} - H_{22} \frac{1}{2} e^{\lambda} - G_{22}\dot\omega e^{-f+\omega} &= - e^{2\omega+\lambda} ~,\\
H_{00} + H_{22}\frac{1}{2} e^{\lambda} + G_{22} \dot\omega e^{-f+\omega} &= e^{2\omega+\lambda} ~,\\
G_{00} - H_{11} (\dot\omega+\dot\lambda)e^{-f+\omega+\lambda}+G_{11}(1-\frac{1}{2}e^{2\lambda}) &=0\label{eq:GHtwo}~,\\
G_{22} + H_{11}\dot\omega e^{-f+\omega} - G_{11}\frac{1}{2} e^{\lambda} &=0~,\\
-G_{00} + H_{22} \dot\omega e^{-f+\omega} - G_{22} \frac{1}{2}e^\lambda &=0~,\\
-G_{11} - H_{22} \dot\omega e^{-f+\omega} +\frac{1}{2} G_{22} e^\lambda &=0~,\\
G_{11}-H_{00} (\dot\omega +\dot\lambda)e^{-f+\omega +\lambda} + G_{00}(1-\frac{1}{2} e^{2\lambda}) &=0 \\\intertext{and}
-G_{22} + H_{00}\dot\omega e^{\omega-f} -\frac{1}{2}G_{00} e^\lambda &=0
\end{align}
\end{subequations}

Rather than attempt to solve these, we make the following observation. First, it is easy to show that $H_{00}=-H_{11}$ and $G_{00}=-G_{11}$. Combining \eqref{eq:GHone} and \eqref{eq:GHtwo} we get
\begin{equation}\label{eq:finaltrick}
\frac{1}{2} H_{11} e^{2\lambda} + 2 H_{11}\left(\dot\omega+\dot\lambda\right)^2 e^{-2 f +2 \omega} =- e^{2\omega +\lambda}~.
\end{equation}
However, $H_{11}$ is a constant and $e^{2\omega}$, $e^{2\lambda}$ and $e^{2 f}$ are rational polynomials of $y$. The left-hand side of \eqref{eq:finaltrick} is then a rational polynomial and so should its right-hand side. However this is true only when $e^\lambda$ is a rational polynomial. This is not true for the NHEK geometry, but it is true for AdS$_4$ in which case $e^\lambda=1$. This concludes what we sought to confirm.

Since the NHEK geometry is the most general solution to the Einstein equations for a metric of such a form \eqref{eq:NHEKmetricExp}, it is difficult to generalize this result. For instance, we can take a metric of the NHEK form so  that it does not satisfy the Einstein equations of motion. We must still impose the condition $R_{\sigma\mu}K^\sigma{}_\nu + R_{\sigma\nu} K^{\sigma}{}_\mu=0$ for a CKY two-form, so that $K$ still maps to a doublet of Killing vectors. We then ask if the image of the map covers $\mathfrak{sl}(2,\mathbb{R})$ or not. Modulo parallel two-forms we arrive at the same set of equations given in \eqref{eq:GHall}. One can then carry on and show that these equations are consistent only if $\lambda=0$ and the space is conformal to AdS$_4$.

\section*{Acknowledgments}
We would like to thank Shoichi Kawamoto for pointing us towards some relevant literature.

\appendix
\section{Conformal Killing-Yano transport}\label{app:cky}
In this section, we rewrite the definition of a conformal Killing-Yano
tensor into a form of parallel transport equation. By the definition of a conformal Killing-Yano $p$-form $K$, there exist a
$\left(p+1\right)$-form $A$ and a $\left(p-1\right)$-form $B$, which satisfy 
\[ \nabla_\mu K\sub{\nu}{1}{p} = A_{\mu}\sub{\nu}{1}{p} +
g_{\mu[\nu_{1}}B\sub{\nu}{2}{p}{}_{]} ~.\]
We thus have that $A_{\mu}\sub{\nu}{1}{p} = \nabla_{[\mu}
K\sub{\nu}{1}{p}{}_{]}$ and
\begin{equation}\label{eq:defB} B\sub{\nu}{2}{p}= \frac{p}{d-(p-1)}
\nabla^\mu K_{\mu}\sub{\nu}{2}{p}~,\end{equation}
where $d$ is the spacetime dimension.
Our aim is to add a $p$-form $C$ and write the definition of $K$ using a covariant derivative $D_\mu O$, where $O$ will
be a section 
$K+A+B+C\in \Lambda^p\oplus \Lambda^{p+1}\oplus\Lambda^{p-1}\oplus \Lambda^p$. 
 
From the $B$ components of the transport equation, we will see that in the case of $p=2$, $B_\mu$ is a Killing
vector for a large class of manifolds, which is an ingredient in the main part. 
A possible application of the full transport equation
is to put it on a computer and search for CKY tensors.

\subsection{A identity}
We begin by writing
\[ \nabla_\mu A_\nu\sub{\nu}{1}{p} - \nabla_\nu A_\mu\sub{\nu}{1}{p} = [\nabla_\mu,\nabla_\nu]K\sub{\nu}{1}{p} - 2 \nabla_{[\mu}g_{\nu][\nu_1} B\sub{\nu}{2}{p}_{]} ~.\]
We do the same with indices exchanged
\begin{align}
\nabla_\nu A_{\mu\nu_1}\sub{\nu}{2}{p} - \nabla_{\nu_1} A_{\mu\nu}\sub{\nu}{2}{p} & = -[\nabla_{\nu},\nabla_{\nu_1}]K_\mu\sub{\nu}{2}{p} +2 \nabla_{[\nu}g_{\nu_1][\mu} B\sub{\nu}{2}{p}_{]} \\
\nabla_{\nu_1} A_{\mu\nu}\sub{\nu}{2}{p} + \nabla_{\mu} A_{\nu}\sub{\nu}{1}{p} & = [\nabla_{\nu_1},\nabla_{\mu}]K_\nu\sub{\nu}{2}{p}-2 \nabla_{[\nu_1}g_{\mu][\nu} B\sub{\nu}{2}{p}_{]}
\end{align}
and add the three equations together to get
\begin{equation}\begin{aligned}
2  \nabla_\mu A_\nu\sub{\nu}{1}{p} &= [\nabla_\mu,\nabla_\nu]K\sub{\nu}{1}{p} -[\nabla_{\nu},\nabla_{\nu_1}]K_\mu\sub{\nu}{2}{p}+[\nabla_{\nu_1},\nabla_{\mu}]K_\nu\sub{\nu}{2}{p}\\&
- 2 \nabla_{[\mu}g_{\nu][\nu_1} B\sub{\nu}{2}{p} _{]}
+2 \nabla_{[\nu}g_{\nu_1][\mu} B\sub{\nu}{2}{p} _{]}
-2 \nabla_{[\nu_1}g_{\mu][\nu} B\sub{\nu}{2}{p}_{]}~.\label{eq:nabla-A}
\end{aligned}\end{equation}

In the following, we often use the identity
\[ X_{[a_1\cdots a_k]} = \frac{1}{k} \left( X_{|a_1| [ a_2\cdots a_k]} - X_{[ a_2 |a_1| a_3\cdots a_k]} + \cdots + (-1)^{k+1} X_{[a_k  a_2 \cdots a_{k-1}] a_1} \right) \]
and any other symmetries of the expression for $X$.
Using them, we write (\ref{eq:nabla-A}) as
\begin{equation}
\begin{aligned}
2  \nabla_\mu A_\nu\sub{\nu}{1}{p} \frac{\grad x^{\nu \nu_1\cdots \nu_p}}{(p+1)!}
&= \Big\{ p R_{\mu\nu\nu_1}{}^\sigma K_{\sigma}\sub{\nu}{2}{p}\\&
- R_{\nu\nu_1\mu}{}^\sigma K_\sigma\sub{\nu}{2}{p} - (p-1) R_{\nu\nu_1\nu_2}{}^\sigma K_{\mu\sigma}\sub{\nu}{3}{p} \\&
+ p R_{\nu_1 \mu \nu}{}^\sigma K_\sigma\sub{\nu}{2}{p} \\&
+ g_{\mu\nu_1}\nabla_\nu B\sub{\nu}2p+\frac{2}p g_{\nu_1\mu}\nabla_\nu B\sub{\nu}{2}{p}-g_{\mu\nu}\nabla_{\nu_1} B\sub{\nu}2p \Big\} \frac{\grad x^{\nu \nu_1\cdots \nu_p}}{(p+1)!}
\end{aligned}
\end{equation}

Using the algebraic Bianchi identity, we collect our first main identity
\[\nabla_\mu A_\nu\sub{\nu}{1}{p} \frac{\grad x^{\nu \nu_1\cdots \nu_p}}{(p+1)!}
=\left(-\frac{p+1}{2} R_{\nu\nu_1\mu}{}^\sigma K_\sigma\sub{\nu}2p - \frac{p+1}{p} g_{\mu\nu} \nabla_{\nu_1} B\sub{\nu}2p \right) \frac{\grad x^{\nu \nu_1\cdots \nu_p}}{(p+1)!}\]
We can also write this as
\[ \nabla_\mu A = -\frac{p+1}{2} R_{\nu\nu_1\mu}{}^\sigma K_\sigma\sub{\nu}2p \frac{\grad x^{\nu \nu_1\cdots \nu_p}}{(p+1)!} - \frac{1}{p^2} g_{\mu\nu} dx^{\nu} \wedge \grad B \]
We also want one more expression for this. It is
\[ \nabla_\mu A_\nu\sub{\nu}1p 
= - R_{\mu\sigma\nu[\nu_1} K^\sigma\sub{\nu}2p_{]} -\frac{p-1}{2} R_{\mu\sigma[\nu_1\nu_2} K^\sigma{}_{|\nu|}\sub{\nu}3p{}_{]} -\frac{1}{p^2} g_{\mu\nu} \grad B\sub{\nu}1p +\frac{1}{p} g_{\mu[\nu_1} \grad B_{|\nu|}\sub{\nu}2p_{]} ~.\]
Using this last expression, we derive
\begin{equation} \nabla_\mu A^\mu\sub{\nu}1p = -
R_{\sigma[\nu_1}K^\sigma\sub{\nu}2p_{]}+\frac{p-1}{2}
R_{\mu\sigma[\nu_1\nu_2}K^{\mu\sigma}\sub{\nu}3p_{]} -\frac{d-p}{p^2}
\grad B\sub{\nu}1p ~.\label{eq:divA}
\end{equation}

\subsection{B identity}
We also need to find an expression for the derivative on $B$. We begin
from \eqref{eq:defB} to obtain
\begin{equation}\begin{aligned}
\nabla_{\nu_1} B\sub{\nu}2p &= \frac{p}{d-(p-1)}\nabla_{\nu_1}\nabla_\mu K^\mu\sub{\nu}{2}{p} \\&=
\frac{p}{d-(p-1)}[\nabla_{\nu_1},\nabla_{\mu}] K^\mu\sub{\nu}{2}{p}+\frac{p}{d-(p-1)} \nabla^\mu \nabla_{\nu_1} K_\mu\sub{\nu}2p \\&=
\frac{p}{d-(p-1)}[\nabla_{\nu_1},\nabla_{\mu}] K^\mu\sub{\nu}{2}{p}+\frac{p}{d-(p-1)} \nabla^\mu \left(-A_{\mu}\sub{\nu}1p+g_{\nu_1[\mu}B\sub{\nu}2p_{]}\right)~.
\end{aligned}\end{equation}
However we also have
\begin{align}
 [\nabla_{\nu_1},\nabla_{\mu}] K^\mu\sub{\nu}{2}{p}
&= R_{\nu_1\mu}{}^{\mu\sigma} K_{\sigma}\sub{\nu}2p + (p-1) R_{\nu_1\mu[\nu_2|\sigma|}K^{\mu\sigma}\sub{\nu}3p_{]} \\
&= - R_{\sigma\nu_1}K^{\sigma}\sub{\nu}2p+\frac{p-1}{2} R_{\mu\sigma\nu_1[\nu_2}K^{\mu\sigma}\sub{\nu}3p_{]}
\end{align}
where we used the algebraic Bianchi identity. Using the identity (\ref{eq:divA}), and putting it all together, we arrive at
\begin{equation}\begin{aligned}
\nabla_{\nu_1} B\sub{\nu}2p &=
 \frac{p}{d-(p-1)} \left( - R_{\sigma\nu_1} K^\sigma\sub\nu{2}p +\frac{p-1}{2} R_{\mu\sigma\nu_1[\nu_2} K^{\mu\sigma}\sub\nu{3}p_{]} \right)\\&
 - \frac{p}{d-(p-1)} \left( - R_{\sigma[\nu_1} K^\sigma\sub\nu{2}p_{]} +\frac{p-1}{2} R_{\mu\sigma[\nu_1\nu_2} K^{\mu\sigma}\sub\nu{3}p_{]} \right) \\&
 +\frac{1}{d-(p-1)} \left( (d-p)\nabla_{[\nu_1}B\sub{\nu}2p_{]}+ \nabla_{\nu_1}B\sub{\nu}2p \right)\\& -\frac{p-1}{d-(p-1)}g_{\nu_1[\nu_2}\nabla_{|\mu|}B^\mu\sub{\nu}3p_{]}~.
\end{aligned}\end{equation}
The trace part with respect to $\nu_1$ and $\nu_2$ gives $\nabla_\mu B^\mu{}\sub{\nu}3p =0$, and collecting terms we get the B identity
\begin{equation}
\begin{aligned}
\nabla_\mu B\sub{\nu}2p &=+ \nabla_{[\mu} B\sub{\nu}2p_{]} + \frac{p}{d-p} \left( R_{\sigma[\mu}K^\sigma\sub{\nu}2p_{]} - R_{\sigma\mu}K^\sigma\sub{\nu}2p \right) \\&
+\frac{p(p-1)}{2(d-p)} \left( R_{\sigma\tau \mu[\nu_2}K^{\sigma\tau}\sub{\nu}3p_{]} - 
R_{\sigma\tau [\mu\nu_2}K^{\sigma\tau}\sub{\nu}3p_{]} \right)
 ~.
\end{aligned}
\end{equation}

\subsection{dB identity}
Define now $C=\grad B$. We have
\begin{multline}
C_{\mu}\subn2p = p \nabla_\mu B\subn2p - 
\frac{p^2}{d-p} \left( 
                      R_{\sigma [\mu} K^\sigma\subn2p_{]} 
                      -R_{\sigma \mu} K^\sigma\subn2p \right)
\\-\frac{{p^2}({p-1})}{2(d-p)}
                \left( R_{\sigma\tau\mu[\nu_2} K^{\sigma\tau}\subn3p_{]} 
                       -R_{\sigma\tau[\mu\nu_2} K^{\sigma\tau}\subn3p_{]} \right)
\end{multline}
Next define
\[ \Omega_{\mu\nu\rho}\subn3p = \nabla_\mu C_{\nu\rho}\subn3p - \nabla_\nu C_{\mu\rho}\subn3p \]
so that \[ 2 \nabla_\mu C_{\nu\rho}\subn3p = \Omega_{\mu\nu\rho}\subn3p + \Omega_{\rho\nu\mu}\subn3p + \Omega_{\rho\mu\nu}\subn3p ~.\]

In calculating $\Omega_{\mu\nu_1\nu_2}\subn3p$, 
we treat the $\nu_2$ index separately from the $\nu_3,\cdots, \nu_p$ indices for
later use. When the $\nabla$ acts
on  $K\subn1p$, we exchange it for  $A$ and $B$ using the CKY equation.
The end formula is
\begin{equation}\begin{aligned}
&\Omega_{\mu\nu_1\nu_2}\subn3p
\stackrel{[\subn3p]}{=}\\&
p [\nabla_\mu,\nabla_{\nu_1}] B\subn2p \\&
+ \frac{p(p-1)}{d-p}\Big( 
      \left(\nabla_\mu R_{\sigma\nu_1}\right) K^\sigma\subn2p - (\mu\leftrightarrow\nu_1)
                    \Big) \\&
          + \frac{p}{d-p}\Big( \left(\nabla_\mu R_{\sigma\nu_2}\right)
		 K^{\sigma}{}_{\nu_1}\subn3p +(p-2)\left(\nabla_{\nu_1}
		 R_{\sigma\nu_3}\right) K^\sigma{}_{\mu\nu_2}\subn4p  - (\mu\leftrightarrow\nu_1)
                    \Big) \\&
+ \frac{p(p-1)}{d-p} R^\sigma_{\nu_1} A_{\mu\sigma}\subn2p - (\mu\leftrightarrow\nu_1) \\&
-\frac{p-1}{d-p} \Big( R^\sigma_{\nu_1} g_{\mu\nu_2} B_\sigma\subn3p + (p-2) R^\sigma_\mu g_{\nu_1\nu_3}B_{\sigma\nu_2}\subn4p   - (\mu\leftrightarrow\nu_1) \Big) \\&
+\frac{p}{d-p}\Big( 
2 R_{\sigma \nu_2}
A_{\mu}{}^\sigma{}_{\nu_1}\subn3p - 2(p-2) R_{\sigma \nu_3} A_\mu{}^\sigma{}_{\nu_1\nu_2}\subn4p 
\Big) \\&
+\frac{1}{d-p} \Big(
R_{\mu\nu_2} B_{\nu_1}\subn3p +(p-2)R_{\nu_1\nu_3}B_{\mu\nu_2}\subn4p + (p-2) R^\sigma_{\nu_2} g_{\mu\nu_3} B_{\sigma\nu_1}\subn4p \\&
+(p-2)R^\sigma_{\nu_3} g_{\nu_1\nu_2} B_{\sigma\mu}\subn4p + (p-2)(p-3) R^\sigma_{\nu_3} g_{\mu\nu_4} B_{\sigma\nu_1\nu_2}\subn5p -  (\mu\leftrightarrow\nu_1) \Big) \\&
- \frac{p(p-2)}{2(d-p)}
\Big( \left(\nabla_\mu R_{\sigma\tau\nu_1\nu_2}\right)
		 K^{\sigma\tau}{}\subn3p +(p-2)\left(\nabla_{\nu_1}
		 R_{\sigma\tau\mu\nu_3}\right) K^{\sigma\tau}{}_{\nu_2}\subn4p \\&
\quad -2\left( \nabla_\mu R_{\sigma\tau\nu_2\nu_3}\right)
		 K^{\sigma\tau}{}_{\nu_1}\subn4p - (p-3)\left(\nabla_\mu
		 R_{\sigma\tau\nu_3\nu_4} \right)
		 K^{\sigma\tau}{}_{\nu_1\nu_2}\subn5p -(\mu\leftrightarrow\nu_1) \Big) \\&
- \frac{p(p-2)}{2(d-p)} \Big( R_{\sigma\tau\nu_1\nu_2}A_{\mu}{}^{\sigma\tau}\subn3p + (p-2) R_{\sigma\tau\mu\nu_3}A_{\nu_1}{}^{\sigma\tau}{}_{\nu_2}\subn4p - (\mu\leftrightarrow\nu_1) \Big) \\&
+ \frac{p(p-2)}{2(d-p)} \Big( 4 R_{\sigma\tau\nu_2\nu_3}A_{\mu}{}^{\sigma\tau}{}_{\nu_1}\subn4p + 2(p-3) R_{\sigma_\tau\nu_3\nu_4}A_{\mu}{}^{\sigma\tau}{}_{\nu_1\nu_2}\subn5p \Big)\\&
- \frac{p-2}{2(d-p)} \Big(  2 R^{\sigma\tau}{}_{\nu_1\nu_2} g_{\mu\sigma}B_{\tau}\subn3p -2(p-2)R^{\sigma\tau}{}_{\nu_1\nu_3} g_{\sigma\mu} B_{\tau\nu_2}\subn4p \\&
\quad +(p-2) R^{\sigma\tau}{}_{\nu_1\nu_2}g_{\mu\nu_3}B_{\sigma\tau}\subn4p - (p-2) R^{\sigma\tau}{}_{\nu_1\nu_3}g_{\mu\nu_2} B_{\sigma\tau}\subn4p \\&
\quad -(p-2)(p-3) R^{\sigma\tau}{}_{\nu_1\nu_4} g_{\mu\nu_3} B_{\sigma\tau\nu_2}\subn5p
-4 R^{\sigma\tau}{}_{\nu_2\nu_3}g_{\mu\sigma} B_{\tau\nu_1}\subn4p \\&
\quad +2(p-3) R^{\sigma\tau}{}_{\nu_4\nu_3}g_{\mu\sigma}B_{\tau\nu_1\nu_2}\subn5p
-2(p-3) R^{\sigma\tau}{}_{\nu_2\nu_3} g_{\mu\nu_4} B_{\sigma\nu_1\tau}\subn5p \\&
\quad +(p-3) R^{\sigma\tau}{}_{\nu_4\nu_3}g_{\mu\nu_2}B_{\sigma\nu_1\tau}\subn5p
+(p-3)(p-4) R^{\sigma\tau}{}_{\nu_5\nu_3} g_{\mu\nu_4} B_{\sigma\nu_1\tau\nu_2}\subn6p\\&
\quad - (\mu\leftrightarrow\nu_1) \Big),
\end{aligned}\end{equation}
where $\stackrel{[\subn3p]}{=}$ indicates that we antisymmetrize the
right-hand side with respect to $\nu_3, \ldots, \nu_p$. 
By alternating indices, and a few pages calculation, we derive a differential condition for $C\subn1p$,
\begin{equation}\begin{aligned}
\nabla_\mu C\subn1p \stackrel{[\subn1p]}{=} &
-\frac{p^2}{2} R_{\nu_1\nu_2\mu}{}^\sigma B_{\sigma}\subn3p \\&
 -\frac{p^2}{d-p} \nabla_{\nu_1} R^\sigma_\mu K_\sigma\subn2p - \frac{p^2}{d-p} \nabla_{\nu_1} R^\sigma_{\nu_2} K_{\sigma\mu}\subn3p \\&
 +\frac{p^2}{d-p} R^{\sigma}_\mu A_{\sigma}\subn1p - \frac{p^2}{d-p} R^{\sigma}_{\nu_1} A_{\mu\sigma}\subn2p \\&
 +\frac{p}{d-p} g_{\mu\nu_1} R^{\sigma}_{\nu_2}B_{\sigma}\subn3p - \frac{p}{d-p} R_{\mu\nu_1} B\subn2p \\&
 +\frac{p^2(p-2)}{4(d-p)} \nabla_\mu R_{\sigma\tau\nu_1\nu_2} K^{\sigma\tau}\subn3p \\&
 +\frac{p^2(p-2)}{2(d-p)} R_{\sigma\tau\nu_1\nu_2} A_\mu{}^{\sigma\tau}\subn3p
 - \frac{p^2(p-2)}{2(d-p)} R_{\sigma\tau\mu\nu_1} A^{\sigma\tau}\subn2p \\&
 + \frac{p(p-2)}{2(d-p)} R_\mu{}^\tau{}_{\nu_1\nu_2} B_{\tau}\subn3p 
 - \frac{p(p-2)}{2(d-p)} g_{\mu\nu_1} R_{\sigma\tau\nu_2\nu_3} B^{\sigma\tau}\subn4p.
\end{aligned}\end{equation}

\subsection{CKY two-forms}
In the case of $p=2$, a Conformal Killing-Yano two-form by definition satisfies
\begin{equation} \nabla_\mu K_{\nu\rho} = A_{\mu\nu\rho} +\frac{1}{2} g_{\mu\nu}B_\rho - \frac{1}{2} g_{\mu\rho}B_\nu \end{equation}
and is in one-to-one correspondence with a section $K\oplus A\oplus B\oplus C$ that is parallel with respect to a connection $D$. This connection is given by the equation above, plus
\begin{align}
\nabla_\mu A_{\nu_1\nu_2\nu_3} &  \stackrel{[\nu_1\nu_2\nu_3]}{=} 
-\frac{3}{2} R_{\nu_1\nu_2\mu}{}^\sigma K_{\sigma\nu_3} - \frac{3}{4} g_{\mu\nu_1} C_{\nu_2\nu_3} \\
\nabla_\mu B_\nu & = \frac{1}{2} C_{\mu\nu} -\frac{1}{d-2} ( R_{\sigma\mu}K^\sigma{}_\nu + R_{\sigma\nu} K^{\sigma}{}_\mu ) \label{eq:nabla-b}\\
\nabla_\mu C_{\nu_1\nu_2} & =
- 2 R_{\nu_1\nu_2\mu}{}^\sigma B_\sigma \notag\\ &
+ \frac{2}{d-2} ( \nabla_{\nu_2} R_{\sigma\nu_1}  K^\sigma{}_\mu - \nabla_{\nu_1} R_{\sigma\nu_2}  K^\sigma{}_\mu + \nabla_{\nu_2}R_{\sigma\mu} K^\sigma{}_{\nu_1} - \nabla_{\nu_1}R_{\sigma\mu} K^\sigma{}_{\nu_2} ) \notag\\&
+\frac{1}{d-2} ( 4 R_{\sigma\mu} A^\sigma{}_{\nu_1\nu_2} + 2 R_{\sigma\nu_1} A^\sigma{}_{\mu\nu_2} - 2 R_{\sigma\nu_2} A^\sigma{}_{\mu\nu_1} )\notag\\&
+ \frac{1}{d-2} ( R_{\sigma\nu_2}B^\sigma g_{\mu\nu_1} - R_{\sigma\nu_1}B^\sigma g_{\mu\nu_2} + R_{\mu\nu_2}B_{\nu_1} - R_{\mu\nu_1}B_{\nu_2}).
\end{align}
Eq. (\ref{eq:nabla-b}) implies that $B_\mu$ is a Killing vector if
$R_{\sigma\mu}K^{\sigma}_{\ \nu}$ is antisymmetric with respect to $\mu$
and $\nu$. This is realized, for instance, if 
\begin{align*}
R_{\mu\nu}=\sigma\left( x\right)g_{\mu\nu}~,
\end{align*}
where $\sigma(x)$ is an arbitrary function.
\section{Massless Kerr is Anti-de Sitter}\label{app:pureAdS}
We give here, for reference, the isometry of the massless, NUT-less, Kerr black hole into AdS$_4$. For simplicity, we rescale here $\hat{t}$ and $\psi$ by  $\ell$ so that the metric becomes
\begin{multline}
   \grad s^2 = \ell^2 \Big\{
\frac{ (r^2+a^2)(r^2+\tilde\ell^2)}{r^2+y^2}\left(  \grad \tilde{t} + y^2  \grad \tilde\psi \right)^2 + 
\frac{r^2+y^2}{ (r^2+a^2)(r^2+\tilde\ell^2)}  \grad r^2 \\ +
\frac{ (y^2-a^2)(y^2-\tilde\ell^2)}{r^2+y^2}\left(  \grad \tilde{t} - r^2  \grad \tilde\psi \right)^2 +
\frac{r^2+y^2}{ (y^2-a^2)(y^2-\tilde\ell^2)} \grad y^2 \Big\}
\end{multline}
with $\tilde\ell=\ell$. It appears that $a/\tilde\ell$ is a physical parameter, while both $a$ and $\tilde{\ell}$ can be scaled with the coordinates freely. However, for $|y|\leq a$, the metric is isometric to the whole of AdS$_4$,
\[  \grad s^2=\ell^2 \left\{ -(1+R^2) \grad T^2 + R^2 \sin^2\Theta^2  \grad \Phi^2 +\frac{ \grad R^2}{1+R^2} + R^2  \grad \Theta^2 \right\} ~,\]
by using the diffeomorphisms
\begin{align}
 \tilde{t} &= \frac{\tilde\ell}{\tilde\ell^2-a^2} T - \frac{a}{\tilde\ell^2-a^2}\Phi \\
 \psi&= \frac{1}{a(\tilde\ell^2-a^2)}\Phi - \frac{1}{\tilde\ell(\tilde\ell^2-a^2)}T\\
 R^2 \sin^2\Theta &= \frac{(r^2+a^2)(a^2-y^2)}{a^2(\tilde\ell^2-{a^2})}\\
1+R^2 &= \frac{(\tilde\ell^2-y^2)(r^2+\tilde\ell^2)}{\tilde\ell^2(\tilde\ell^2-a^2)}~.
\end{align}
For $|y|>\tilde\ell$ we simply exchange $a$ for $\tilde\ell$. However in this case, the diffeomorphism covers only half of the two-sphere $\cos\Theta>0$.

\section{Profiles in de Sitter Kerr}\label{app:dSProfiles}
We give here a derivation of the profiles of $\Delta_y$ and $\Delta_r$ for positive cosmological constant. These results supplement the numerics of \cite{Akcay:2010vt}. As with negative cosmological constant, our tool is the deformation of their graphs by varying $M$ and $L$.

For $M=0$ there are always two roots and one bounded region where $\Delta_r$ is positive. Whether there is one or three extrema depends, respectively, on $1<a^2 g^2$ or $1\geq a^2 g^2$. As we turn on $M$, there will be a change in the number of roots of $\Delta_r$ when an extremum of $\Delta_r$ crosses the horizontal axis: $\Delta_r=\Delta_r'=0$. Eliminating $M$ from the two equations, gives
\[ 3 g^2 r^4 - (1-a^2 g^2)r^2+ a^2 = 0~,\]
with solutions of positive $r^2$ only when $1-a^2 g^2>0$. The two solution are
\[ \bar{r}_\pm^2 =\frac{1}{6g^2} \left( 1-a^2 g^2 \pm \sqrt{ (1-a^2g^2)^2-12 a^2 g^2}\right) ~,\]
and the determinant is non-negative when $|a^2 g^2 - 7|\geq 4\sqrt{3}$. So there is no change in roots, unless $a^2g^2\leq 7-2\sqrt{12}$, in which case there are two transitions at $M_-$ and $M_+$.

\begin{figure}
\centering
{\includegraphics[width=0.5\textwidth]{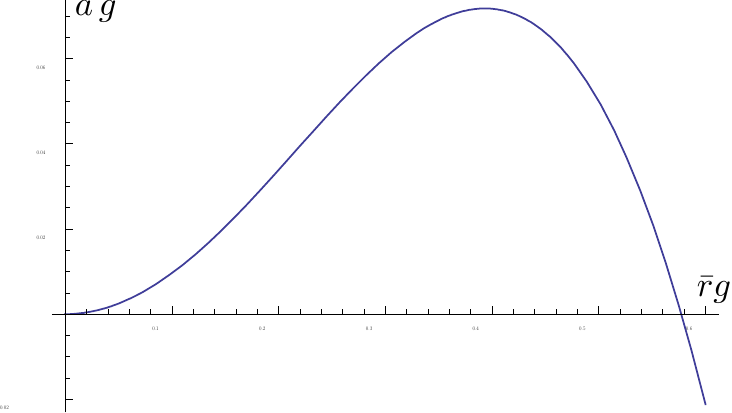}}   
  \quad        
    \caption{Extremal $\Delta_r$, $a$ as a function of $\bar{r}$.}
  \label{fig:agVsbarrdS}
\end{figure}

\begin{figure}
  \centering
  \subfloat[always two roots]{\includegraphics[width=0.3\textwidth]{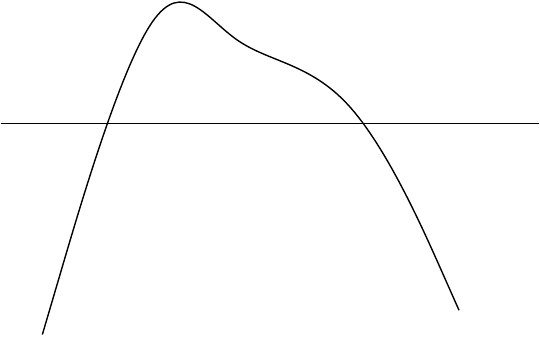}}   
  \quad       
  \subfloat[$M<M_-$]{\includegraphics[width=0.3\textwidth]{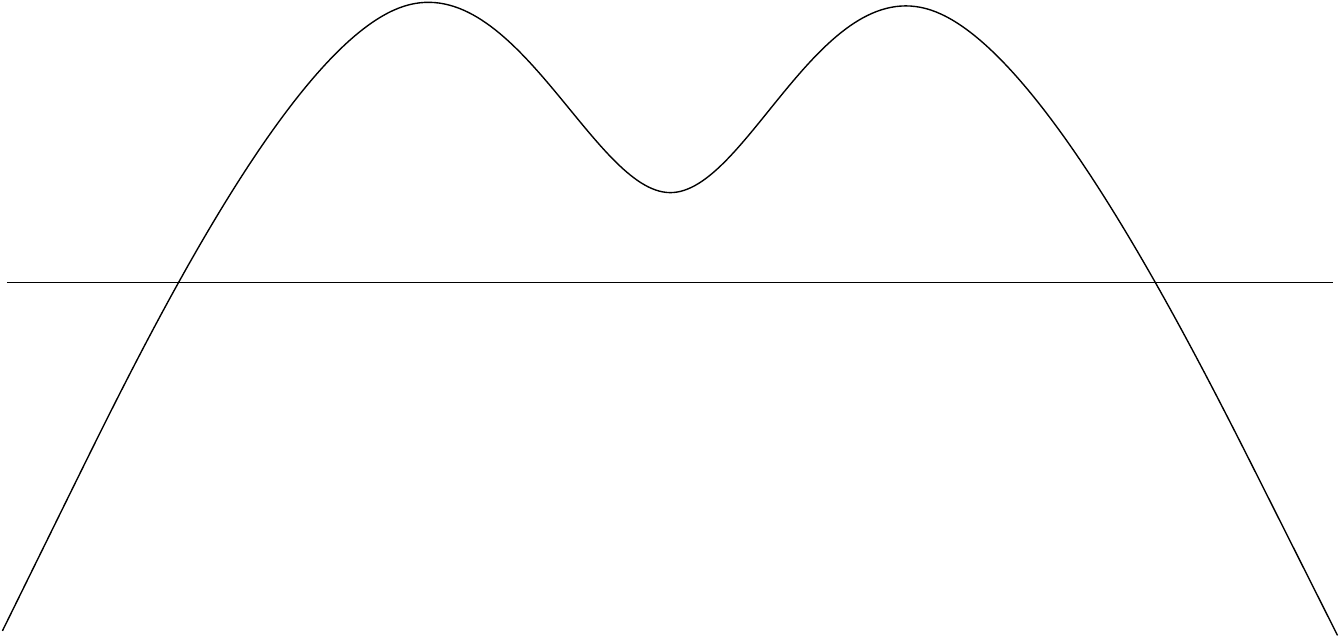}} 
  \\
  \subfloat[$M_-<M<M_+$]{\includegraphics[width=0.3\textwidth]{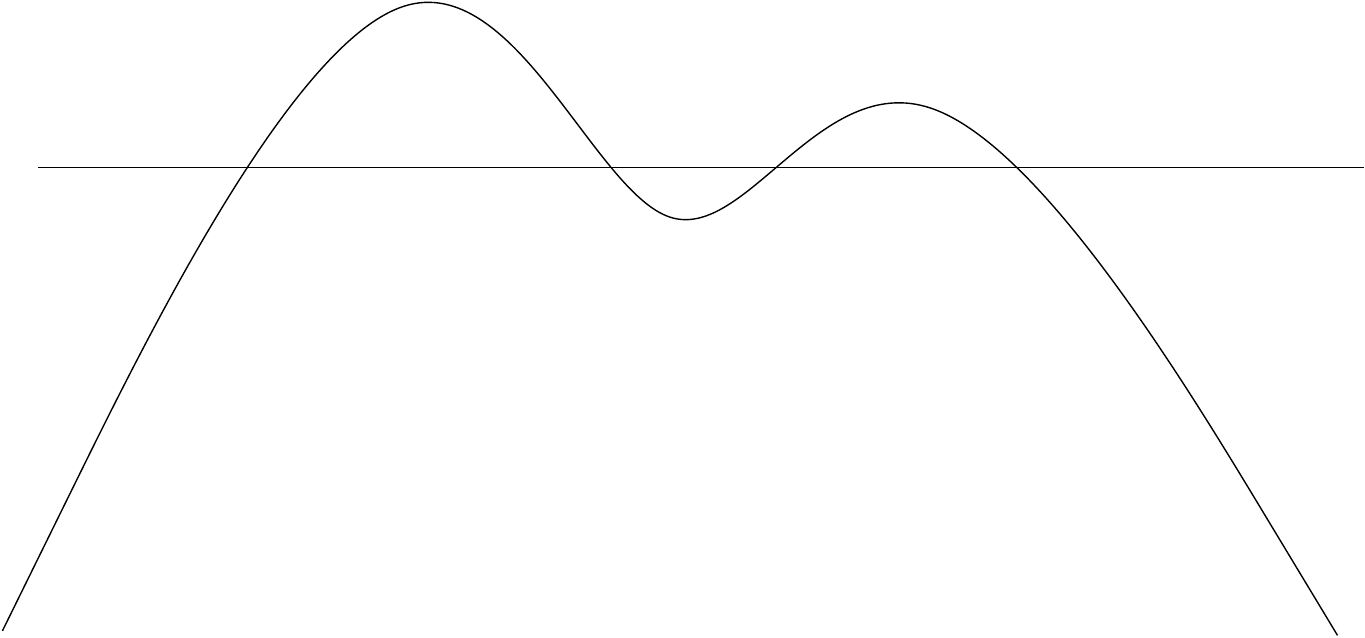}}   \quad
  \subfloat[$M>M_+$]{\includegraphics[width=0.3\textwidth]{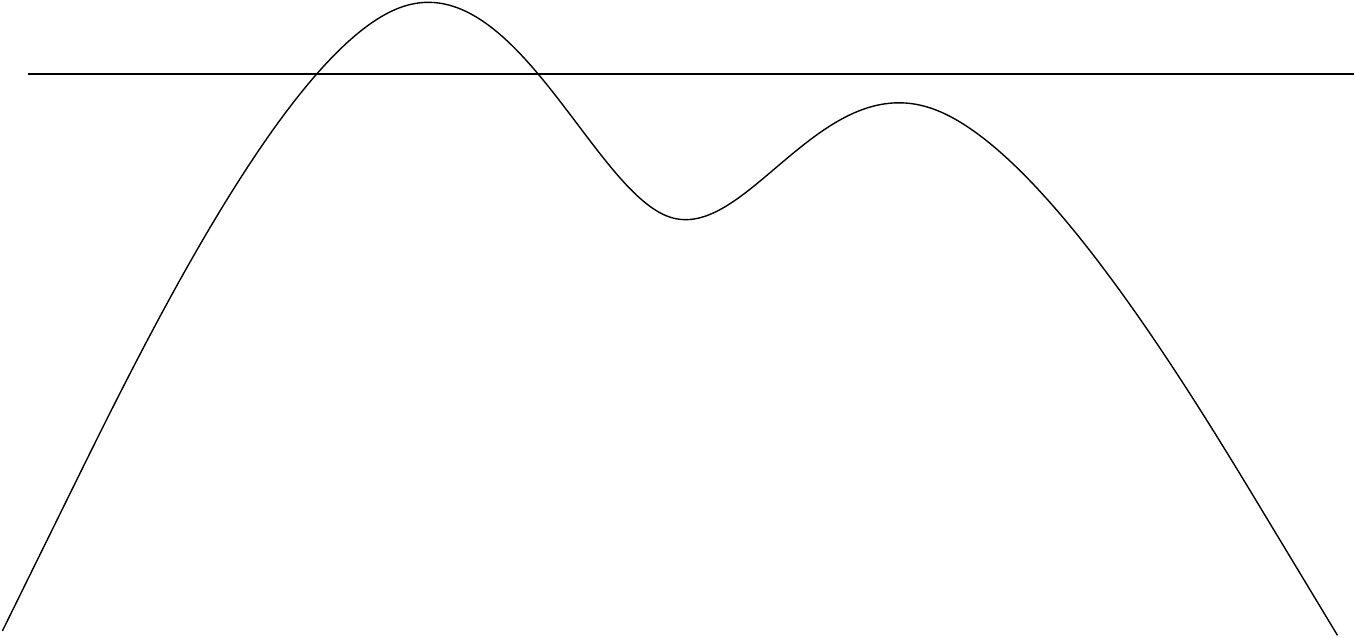}}   
  \caption{Graph of $\Delta_r(r)$ in Kerr-dS for fixed $a$. The graph in (a) is for the case $a^2 g^2>7-4\sqrt{3}$ and the graphs in (b)-(d) are for $a^2 g^2<7-4\sqrt{3}$.}
  \label{fig:TransDeltaRdS}
\end{figure}

The extremal parameters $(a,M)$ are parametrized in terms of $\bar{r}$ as
\begin{align}
M   &=   \bar{r} \frac{ \left( 1- g^2\bar{r}^2 \right)^2 }{1+ g^2\bar{r}^2 } &
a^2 &=   \bar{r}^2 \frac{  1- 3g^2\bar{r}^2  }{1+g^2\bar{r}^2 }~.
\end{align}
However, the function $\bar{r}\mapsto a^2$ is not one-to-one, see figure~\ref{fig:agVsbarrdS}. That is, for any $a$ there are two critical values of $M$, $M_\pm$, where $\Delta_r$ has a double root.  The profile of $\Delta_r$ is shown in figure~\ref{fig:TransDeltaRdS}. The parametric plot of $(M^2,a)$ as a function of $\bar{r}$ was drawn in figure~\ref{fig:physMdS}.

\section{AdS3 coordinates}\label{app:AdS3}
Anti de-Sitter space in $d=3$ has isometry algebra
\begin{equation}
\mathfrak{so}(2,2) = \mathfrak{sl}(2,\mathbb{R})_{\mathrm{L}}\oplus \mathfrak{sl}(2,\mathbb{R})_{\mathrm{R}} = \langle l_a \rangle_{a=0,1,2} \oplus \langle r_a \rangle_{a=0,1,2}~,
\end{equation}
where we choose a basis such that $[r_a,r_b]=-\epsilon_{ab}{}^c r_c$ and $[l_a,l_b]=-\epsilon_{ab}{}^c l_c$. 

The (universal cover of) AdS$_3$ metric
 $ds^2_{\AdS}$ can be described in coordinates $(x,u,\tau)$ so that $\partial_u$ and $\partial_\tau$ are manifest commuting isometries and $x$ is hypersurface orthogonal. Up to isometries, diffeomorphisms $x\mapsto x'(x)$, parity transformations, and $\mathrm{GL}(2,\mathbb{R})$ matrix transformations on $(u,\tau)$, the classification of $\mathfrak{sl}(2,\mathbb{R})$ elements results to the following choices:
\begin{enumerate}
	\item Global coordinates, $\partial_\tau=\frac{1}{2}(r_0+l_0)$ and $\partial_\phi=\frac{1}{2}(r_0-l_0)$
	\begin{equation}
	 \grad s^2_{\AdS} =  -(1+x^2) \grad \tau^2 + \frac{ \grad x^2}{x^2+1} + x^2  \grad  u^2~,
  \end{equation}
  with $u=u+2\pi$, $\tau\in\mathbb{R}$ and $x\geq 0$.
  \item Spacelike self-dual global coordinates, for which $\partial_u = l_2$ and $\partial_\tau=r_0$
  \begin{equation}
	 \grad s^2_{\AdS} =  \frac{1}{4}\left( -(x^2+1) \grad \tau^2 + \frac{ \grad x^2}{x^2+1} + \left( \grad u+x\, \grad \tau\right)^2\right)~,
  \end{equation}
  with $x,u,\tau\in\mathbb{R}$. It covers the space globally.
  \item Spacelike self-dual non-extremal coordinates, for which $\partial_u = l_2$ and $\partial_\tau=r_2$
  \begin{equation}
	\grad s^2_{\AdS} =  \frac{1}{4}\left( -(x^2-1) \grad \tau^2 + \frac{ \grad x^2}{x^2-1} + \left( \grad u+x\, \grad \tau\right)^2\right)~,
  \end{equation}
  with $x,u,\tau\in\mathbb{R}$.
  \item Spacelike self-dual extremal coordinates, for which $\partial_u = l_2$ and $\partial_\tau=r_0+r_2$
  \begin{equation}\label{eq:AdSSD}
	\grad s^2_{\AdS} =  \frac{1}{4}\left( -x^2\, \grad \tau^2 + \frac{ \grad x^2}{x^2} + \left( \grad u+x\, \grad \tau\right)^2\right)~,
  \end{equation}
  with $x,u,\tau\in\mathbb{R}$. These are the relevant coordinates we used in the NHEK.
  \item Timelike self-dual coordinates, for which $\partial_\tau=r_0+l_2$ and $\partial_u= l_0$,
  \begin{equation}
  \grad s^2_{\AdS} =  \frac{1}{4}\left( x^2\, \grad \tau^2 + \frac{ \grad x^2}{x^2} - \left( \grad u+x\, \grad \tau\right)^2\right)~,
  \end{equation}
  with $x,u,\tau\in\mathbb{R}$. These are the relevant coordinates we used in the polar limit.
  \item
  Poincar\'e coordinates, for which $\partial_\tau=r_0+r_2$ and $\partial_u=l_0+ l_2$,
  \begin{equation}
  \grad s^2_{\AdS} =   \frac{ \grad x^2}{x^2} + x^2  \grad u\, \grad \tau ~,
  \end{equation}
  with $u,\tau\in\mathbb{R}$ and $x>0$.
\end{enumerate}
In the above we have set the cosmological length $R=1$.

A parametrization of the quadric
\[
\left(X^{-1}\right)^{2}+\left(X^{0}\right)^{2}-\left(X^{1}\right)^{2}-\left(X^{2}\right)^{2}=R^{2}.
\]
in terms of the extremal spacelike self-dual coordinates, $b=0$, can be achieved by ($-\infty<\tau<\infty,-\infty<u<\infty,0<x$)
\begin{align*}
A^{+}\equiv X^{-1}+X^{1} & =R\sqrt{x}\sinh\frac{u}{2}\\
A^{-}\equiv X^{-1}-X^{1} & =-R\left(\tau\sqrt{x}\cosh\frac{u}{2}+\frac{1}{\sqrt{x}}\sinh\frac{u}{2}\right)\\
B^{+}\equiv X^{0}+X^{2} & =R\left(\tau\sqrt{x}\sinh\frac{u}{2}+\frac{1}{\sqrt{x}}\cosh\frac{u}{2}\right)\\
B^{-}\equiv X^{0}-X^{2} & =R\sqrt{x}\cosh\frac{u}{2}\ ,
\end{align*}
and the 4 dimensional metric $g_{MN}=diag\left(-1,-1,+1,+1\right)$.
Note that this parametrization covers only the region with $B^{+}>0$
and $B^{+}>A^{+}$. A parametrization of the global spacelike  self-dual
coordinates, $b=1$ in \eqref{eq:AdSSD}, was given in
\cite{Coussaert:1994tu}. This can be related to the above embedding
after an infinite boost $b\rightarrow 0$, see for instance
\cite{Jugeau:2010nq}. 
The right invariant 1-forms $\theta^a$ and the
left invariant 1-forms $\hat{\theta}^a$ are expressed as
\begin{align*}
\theta^{0} &
 =-\frac{2}{R^{2}}\left[X^{-1} \grad X^{0}-X^{0} \grad X^{-1}+X^{1} \grad X^{2}-X^{2} \grad X^{1}\right]\\
\theta^{1} &
 =-\frac{2}{R^{2}}\left[X^{-1} \grad X^{1}-X^{1} \grad X^{-1}+X^{0} \grad X^{2}-X^{2} \grad X^{0}\right]\\
\theta^{2} & =-\frac{2}{R^{2}}\left[X^{-1} \grad X^{2}-X^{2} \grad X^{-1}-X^{0} \grad X^{1}+X^{1} \grad X^{0}\right]
\\
\tilde{\theta}^{0} &
 =\frac{2}{R^{2}}\left[X^{-1} \grad X^{0}-X^{0} \grad X^{-1}-X^{1} \grad X^{2}+X^{2} \grad X^{1}\right]\\
\tilde{\theta}^{1} &
 =\frac{2}{R^{2}}\left[X^{-1} \grad X^{1}-X^{1} \grad X^{-1}-X^{0} \grad X^{2}+X^{2} \grad X^{0}\right]\\
\tilde{\theta}^{2} &
 =\frac{2}{R^{2}}\left[X^{-1} \grad X^{2}-X^{2} \grad X^{-1}+X^{0} \grad X^{1}-X^{1} \grad X^{0}\right]\ ,
\end{align*}
regardless of what metric we choose in 2+2 dimensions.

\section{Einstein solutions of NHEK-type}\label{app:solvingNHEK}
Let us assume a metric of NHEK-type
\begin{equation}
 \grad s^{2}=e^{2\omega\left(y\right)}\left[ -x^2\, \grad \tau^2 + \frac{ \grad x^2}{x^2} + e^{2\lambda\left(y\right)} \left( \grad u+x\, \grad \tau\right)^2 \right]+e^{2f\left(y\right)} \grad y^2.\label{eq:metric-ansatz}
\end{equation}
We can make use of the computation of the curvature tensor in \S\ref{sec:noparallel} in order to derive its Ricci tensor. The non-trivial components of the Einstein equation $R_{AB}=-\frac{3}{\ell^{2}}\eta_{AB}$ come from the diagonal flat components, which are
\begin{align}
e^{-2\omega}-\frac{1}{2}e^{-2\omega+2\lambda}+\left(\ddot{\omega}+3\dot{\omega}^{2}+\dot{\omega}\left(\dot{\lambda}-\dot{f}\right)\right)e^{-2f} & =\frac{3}{\ell^{2}}\\
-e^{-2\omega}+\frac{1}{2}e^{-2\omega+2\lambda}-\left(\ddot{\omega}+3\dot{\omega}^{2}+\dot{\omega}\left(\dot{\lambda}-\dot{f}\right)\right)e^{-2f} & =-\frac{3}{\ell^{2}}\\
-\frac{1}{2}e^{-2\omega+2\lambda}-\left(\ddot{\omega}+3\dot{\omega}^{2}+\dot{\omega}\left(4\dot{\lambda}-\dot{f}\right)+\ddot{\lambda}+\dot{\lambda}^{2}-\dot{\lambda}\dot{f}\right)e^{-2f} & =-\frac{3}{\ell^{2}}\\
-\left(3\ddot{\omega}+3\dot{\omega}^{2}+\dot{\omega}\left(2\dot{\lambda}-3\dot{f}\right)+\ddot{\lambda}+\dot{\lambda}^{2}-\dot{\lambda}\dot{f}\right)e^{-2f} & =-\frac{3}{\ell^{2}}~.
\end{align}
The first and the second equations are the same. A further
simplification is made by taking the gauge $f\left(y\right)=0$. Thus
the equations of motion are reduced to the following three
\begin{align}
e^{-2\omega}-\frac{1}{2}e^{-2\omega+2\lambda}+\left(\ddot{\omega}+3\dot{\omega}^{2}+\dot{\omega}\dot{\lambda}\right) & =\frac{3}{\ell^{2}}\label{eq:eom1}\\
\frac{1}{2}e^{-2\omega+2\lambda}+\left(\ddot{\omega}+3\dot{\omega}^{2}+4\dot{\omega}\dot{\lambda}+\ddot{\lambda}+\dot{\lambda}^{2}\right) & =\frac{3}{\ell^{2}}\label{eq:eom2}\\
3\ddot{\omega}+3\dot{\omega}^{2}+2\dot{\omega}\dot{\lambda}+\ddot{\lambda}+\dot{\lambda}^{2} & =\frac{3}{\ell^{2}}\nonumber ~.
\end{align}

The triplet $(\omega,\dot\omega,\ddot\omega)$ can be algebraically solved in terms of $\lambda$ and its derivatives,
\begin{align}
 \omega&=f_1(\lambda,\dot\lambda,\ddot\lambda)~,\\
\dot\omega&=f_2(\lambda,\dot\lambda,\ddot\lambda)~,\\
\ddot\omega&=f_3(\lambda,\dot\lambda,\ddot\lambda)~.
\end{align}
Since $\dot\omega$ appears quadratically, there are two branches to this solution. However, the gauge $f=0$ is preserved by $y\mapsto-y$ and in fact the two branches are seen to be equivalent. Whence $\omega$ is given in terms of $\lambda$ and its derivatives, provided that the solution for $\lambda$ is consistent with the above three equations. That is, we have $\dot{f}_1=f_2$ and $\dot{f}_2=f_3$. By computer algebra we find that these two differential equations are identical.

At this point the solution to \eqref{eq:metric-ansatz} is given by a third-order highly non-linear differential equation $\dot{\ddot\lambda}=f(\lambda,\dot\lambda,\ddot\lambda)$. It is difficult to solve, even more so due to the gauge we chose. We make the observation that the gauge $f=0$ is also preserved by $y\mapsto y+b$ for any constant $b$ and thus the differential equation has only two rather than three gauge-invariant degrees of freedom. This is the same number of parameters as for the NHEK geometry, $\bar{r}^2$ and $L$. Therefore, up to subtleties on the range of the parameters, the NHEK solution is the most generic Einstein solution to \eqref{eq:metric-ansatz}.

\bibliography{ckyV2}
\bibliographystyle{utphys}
\end{document}